\documentclass[acsmall, review=false, screen=true]{acmart}

\usepackage{hyperref}

\usepackage{annotations}
\usepackage{subfigure}
\usepackage{booktabs}
\usepackage{url}
\usepackage{rotating}
\usepackage{multirow}
\usepackage{tabularx}
\usepackage{xcolor}
\usepackage{framed}
\usepackage{amssymb,amsmath}
\usepackage{ifxetex,ifluatex}
\usepackage{fancyvrb}
\usepackage{enumerate}
\usepackage{amsfonts}
\usepackage{mathtools}

\usepackage{microtype}

\usepackage{siunitx}
\usepackage{xspace}

\usepackage{amsthm}

\def\imagetop#1{\vtop{\null\hbox{#1}}}

\newcommand{\toolname}[0]{{\small\textup{COSMOS}}\xspace}
 
\newcommand{\fillparagraph}{\unskip\parfillskip 0pt \par}
 
\hypersetup{%
pdfauthor   = {Luca Piccolboni, Paolo Mantovani, Giuseppe Di Guglielmo, Luca P. Carloni},
pdfsubject  = {ACM Transactions on Embedded Computing Systems (TECS), CODES+ISSS'17},
pdftitle    = {COSMOS: Coordination of High-Level Synthesis and Memory Optimization for Hardware Accelerators},
bookmarks   = true,
colorlinks  = true,
linkcolor   = black,
citecolor   = black,
urlcolor    = black,
}%

\makeatletter
\newcommand{\removelatexerror}{\let\@latex@error\@gobble}
\makeatother

\newcommand{\accsmall}[1]{{\small\textsc{#1}}}
\newcommand{\accfoot}[1]{{\footnotesize\textsc{#1}}}

\usepackage{adjustbox}

\usepackage{tikz}
\usepackage{pgfplots}
\usepackage{pgfplotstable}
\usetikzlibrary{arrows, shapes, snakes, automata, backgrounds, calc, petri}
\usetikzlibrary{patterns}

\theoremstyle{problem}
\newtheorem{problem}{Problem}

\theoremstyle{plain}
\newtheorem{ex}{Example}

\usepackage[vlined]{algorithm2e}
\usepackage{balance}

\let\oldnl\nl
\newcommand{\nonl}{\renewcommand{\nl}{\let\nl\oldnl}}

\makeatletter
\newcommand{\algrule}[1][.8pt]{\par\vskip.3\baselineskip\hrule width 0.945\textwidth height #1\par\vskip.25\baselineskip}
\makeatother

\SetCommentSty{mycommfont}
\SetNlSty{mynumfont}{}{}

\acmJournal{TECS}

\setcopyright{acmlicensed}

\acmDOI{0000001.0000001}

\received{April 2017}
\received[revised]{May 2017}
\received[accepted]{June 2017}

\acmJournal{TECS}
\acmYear{2017}
\copyrightyear{2017}

\begin{document}

%
%

\title[COSMOS: Coordination of High-Level Synthesis and Memory Optimization for Hardware Accelerators]
{COSMOS: Coordination of High-Level Synthesis \\ and Memory Optimization for Hardware Accelerators}

%
%

\author{Luca Piccolboni}
\orcid{0000-0003-0094-4960}
\affiliation{\institution{Columbia University}
\department{Computer Science}\city{New York}\state{USA}}
\email{piccolboni@cs.columbia.edu}

\author{Paolo Mantovani}
\affiliation{\institution{Columbia University}
\department{Computer Science}\city{New York}\state{USA}}
\email{paolo@cs.columbia.edu}

\author{Giuseppe Di Guglielmo}
\affiliation{\institution{Columbia University}
\department{Computer Science}\city{New York}\state{USA}}
\email{giuseppe@cs.columbia.edu}

\author{Luca P. Carloni}
\affiliation{\institution{Columbia University}
\department{Computer Science}\city{New York}\state{USA}}
\email{luca@cs.columbia.edu}


%
%

\begin{abstract}%
Hardware accelerators are key to the efficiency and performance of
system-on-chip (SoC) architectures. With high-level synthesis (HLS), designers
can easily obtain several performance-cost trade-off implementations for each
component of a complex hardware accelerator. However, navigating this design
space in search of the Pareto-optimal implementations at the system level is a
hard optimization task. We present \toolname, an automatic methodology for the
design-space exploration (DSE) of complex accelerators, that coordinates both
HLS and memory optimization tools in a compositional way. First, thanks to the
co-design of datapath and memory, \toolname produces a large set of
Pareto-optimal implementations for each component of the accelerator. Then,
\toolname leverages compositional design techniques to quickly converge to the
desired trade-off point between cost and performance at the system level. When
applied to the system-level design (SLD) of an accelerator for wide-area motion
imagery (WAMI), \toolname explores the design space as completely as an
exhaustive search, but it reduces the number of invocations to the HLS tool
by up to 14.6$\times$.
\fillparagraph 
\end{abstract}%

%
%



\thanks{The authors are within the Department of Computer Science, Columbia University, New York, NY, USA 
(Luca Piccolboni: piccolboni@cs.columbia.edu, Paolo Mantovani: paolo@cs.columbia.edu, Giuseppe Di
Guglielmo: giuseppe@cs.columbia.edu, and Luca P. Carloni: luca@cs.columbia.edu). \\
This article was presented in the International Conference on Hardware/Software Codesign
and System Synthesis (CODES+ISSS) 2017 and appears as part of the ESWEEK-TECS special issue.}

\maketitle


%
%

\section{Introduction}\label{section:introduction}

{
High-performance systems-on-chip (SoCs) are increasingly based on heterogeneous
architectures that combine general-purpose processor cores and specialized
hardware accelerators~\cite{borkar11,Horowitz2014,carloni_dac16}. Accelerators
are hardware devices designed to perform specific functions.  Accelerators are
become popular because they guarantee considerable gains in both performance
and energy efficiency with respect to the corresponding software
executions~\cite{chen_14, cong_dac14, zhang_15, liu_fpga2016, reagen2016,
kim_tnn17, chen_17, ham_micro16}.  However, the integration of several
specialized hardware blocks into a complex accelerator is a difficult design
and verification task. In response to this challenge, we advocate the
application of two key principles. First, to cope with the increasing
complexity of SoCs and accelerators, most of the design effort should move away
from the familiar register-transfer level (RTL) by embracing \emph{system-level
design} (SLD)~\cite{vincentelli_ieee07,GHP+09} {with high-level synthesis
(HLS)~\cite{meeus_12, qamar_17}.} Second, it is necessary to create reusable
and flexible components, also known as \emph{intellectual property} (IP)
blocks, which can be easily (re)used across a variety of architectures with
different targets for performance and metrics for cost.
\fillparagraph 
}

\subsection{System-Level Design}

{ 
SLD has been proposed as a viable approach to cope with the increasing
complexity of today architectures~\cite{vincentelli_ieee07,GHP+09}.  The SoC
complexity is growing as a result of integrating a larger number of
heterogeneous accelerators on the same chip. Further, accelerators are
themselves becoming more complex to meet the high-performance and low-power
requirements of emerging applications, e.g, deep-learning
applications~\cite{zhang_15, reagen2016, kim_tnn17, chen_17}.  To address the
complexity of systems and accelerators, SLD aims at raising the level of
abstraction of hardware design by replacing cycle-accurate low-level
specifications (i.e., RTL Verilog or VHDL code) with untimed or
transaction-based high-level specifications (i.e., C, C++ or SystemC
code)~\cite{qamar_17}. This allows designers to focus on the relations between
the data structures and computational kernels that characterize the
accelerators, quickly evaluate different alternative implementations of the
accelerators, and perform more complex and meaningful full-system simulations
of the entire SoC.  Indeed, designers can ignore low-level logic and circuit
details that burden the design process. This improves the productivity and
reduces the chances of errors~\cite{carloni_dac16}. 
\fillparagraph
}

Unfortunately, current HLS tools are not ready yet to handle the complexity of
today accelerators. Many accelerators are too complex to be synthesized by
state-of-the-art HLS tools without being partitioned first.  Accelerators must
be decomposed into several computational blocks, or \emph{components}, to be
synthesized and explored efficiently.  {Decomposing an accelerator also helps
improve the quality of results}.  Indeed, the choice of a particular RTL
implementation for a component must be made in the context of the choices for
all the other accelerator components. A particular set of choices leads to one
point in the multi-objective design space of the accelerator. Thus, the process
of deriving the diagram of Pareto-optimal points repeats itself hierarchically
from the single component to the entire accelerator. {This complexity is not
handled by current HLS tools that {optimize the single components independently
from the others.}}

\subsection{Intellectual Property Reuse}

{
HLS supports IP block reuse and exchange.  For instance, a team of
computer-vision experts can devise an innovative algorithm for object
recognition, design a specialized accelerator for this algorithm with a
high-level language (C, C++, SystemC), and license it as a synthesizable IP
block to different system architects; the architects can then exploit HLS tools
to derive automatically the particular implementation that provides the best
trade-off point (e.g., higher performance or lower area/power) for their
particular system. The main idea of HLS is to raise the abstraction level of
the design process to allow designers to generate multiple RTL implementations
that can be reused across many different architectures. To obtain such a
variety of implementations, the designers can change high-level configuration
options, known as \emph{knobs}, so that HLS can transform automatically the
high-level specification of the accelerator and obtain several RTL
implementations with different performance figures and implementation costs.
For example, {loop unrolling} is a knob that allows designers to replicate
parts of the logic to distribute computation in space (resource replication),
rather than in time.  The application of this knob generally leads to a faster,
but larger, implementation of the initial specification. 
\fillparagraph
}

{
Despite the advantages of HLS, performing this \emph{design-space exploration}
(DSE) is still a complicated task, especially for complex hardware
accelerators. {First, the support for memory generation and optimization is
limited {in current HLS tools}. Some HLS tools still require third-party
generators to provide a description of the memory organization and automatize
the DSE process~\cite{pilato_codes14, pilato_tcad17}.} Several studies,
however, highlight the importance of \emph{private memories} to sustain the
parallel datapath of accelerators: on a typical accelerator design, memory
takes from 40\% to 90\% of the area~\cite{Lyons2012,cota_dac15}; hence, its
optimization cannot be an independent task.  Second, HLS tools are based on
{heuristics}, whose behavior is not robust and often hard to
\emph{predict}~\cite{panda07}.  Small changes to the knobs, e.g., changing the
number of iterations unrolled in a loop, can cause significant and unexpected
modifications at the implementation level.  This increases the DSE effort
because small changes to the knobs can take the exploration far {from the
Pareto-optimality.}
\fillparagraph
}

\begin{figure*}[t]
\centering\resizebox{\textwidth}{!}{
\includegraphics[trim={0.4cm 0 0.4cm 0}]{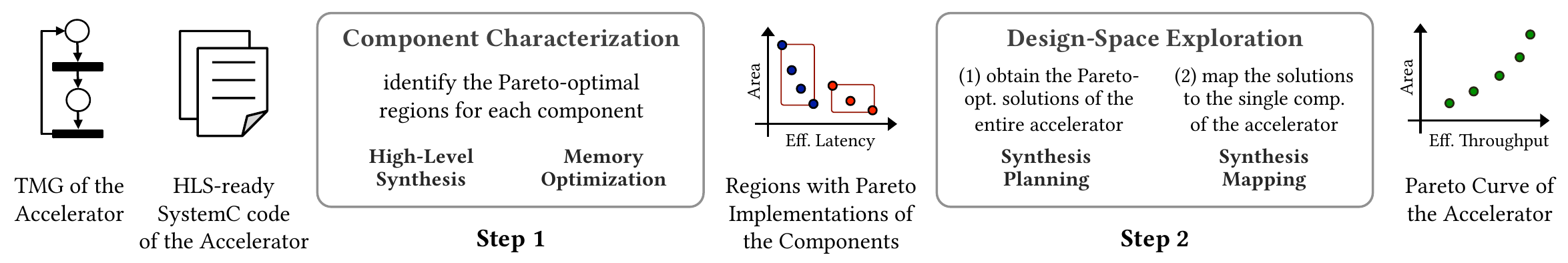}}
\caption{{COSMOS: a methodology to coordinate HLS and memory optimization 
for the DSE of hardware accelerators.}}\label{figure:overview}
\end{figure*}

\subsection{Contributions}

{To address these limitations, we present \toolname\footnote{COSMOS stands for
\emph{``COordination of high-level Synthesis and Memory Optimization for
hardware acceleratorS''}.  We also adopt the name COSMOS for our methodology
since it is the opposite of CHAOS (in the Greek creation myths).  In our
analogy, CHAOS corresponds to the complexity of the DSE process.  }: an
automatic methodology for the DSE of complex hardware accelerators, which are
composed of several components. {\toolname is based on a \emph{compositional}
approach that coordinates both HLS tools and memory generators. First, thanks
to the datapath and memory co-design, \toolname produces a large set of
Pareto-optimal implementations for each component, thus increasing both
performance and cost spans. These spans are defined as the ratios between the
maximum value and the minimum value for performance and cost, respectively.
Second, \toolname leverages compositional design techniques to significantly
reduce the number of invocations to the HLS tool and the memory generator.} In
this way, \toolname focuses on the most critical components of the accelerator
and quickly converges to the desired trade-off point between cost and
performance for the entire accelerator.  The \toolname methodology consists of
two main steps (\figurename~\ref{figure:overview}).  First, \toolname uses an
algorithm to {characterize} each component of the accelerator individually by
efficiently coordinating multiple runs of the HLS and memory generator tools.
This algorithm finds the regions in the design space of the components that
include the Pareto-optimal implementations (\emph{Component Characterization}
in \figurename~\ref{figure:overview}). Second, \toolname performs a DSE to
identify the Pareto-optimal solutions for the entire accelerator by efficiently
solving a linear \mbox{programming (LP) problem instance (\emph{Design-Space
Exploration}).} 
\fillparagraph
}

%
%


We evaluate the effectiveness and efficiency of the \toolname methodology on a
complex accelerator for wide-area motion imagery (WAMI)~\cite{porter10,
perfect-suite-man}, which consists of approximately 7000 lines of SystemC code.
While exploring the design space of WAMI, \toolname returns an average
performance span of 4.1$\times$ and an average area span of 2.6$\times$, as
opposed to 1.7$\times$ and 1.2$\times$ when memory optimization is not
considered and only standard dual-port memories are used.  Further, \toolname
achieves the target data-processing throughput for the WAMI accelerator while
reducing the number of invocations to the HLS tool per component by up to
14.6$\times$, with respect to an exhaustive exploration approach.

\subsection{Organization} 

{The paper is organized as follows.  Section~\ref{section:background} provides
the necessary background for the rest of the paper.
Section~\ref{section:example} describes few examples to show the effort
required in the DSE process.  Section~\ref{section:design} gives an overview
of the \toolname methodology, which is then detailed in Sections
\ref{section:compdse} (\emph{Component Characterization}) and
\ref{section:system} (\emph{Design-Space Exploration}).
Section~\ref{section:results} presents the experimental results.
Section~\ref{section:related} discusses the related work. Finally,
Section~\ref{section:concl} concludes the paper.}


%
%

\section{Preliminaries}\label{section:background}

This section provides the necessary background concepts. We first describe the
main characteristics of the accelerators targeted by \toolname in Section
~\ref{section:background:acc}. Then, we present the computational model we
adopt for the DSE in Section~\ref{section:background:tmg}.

\begin{figure}
\centering\resizebox{0.6\linewidth}{!}{
\includegraphics{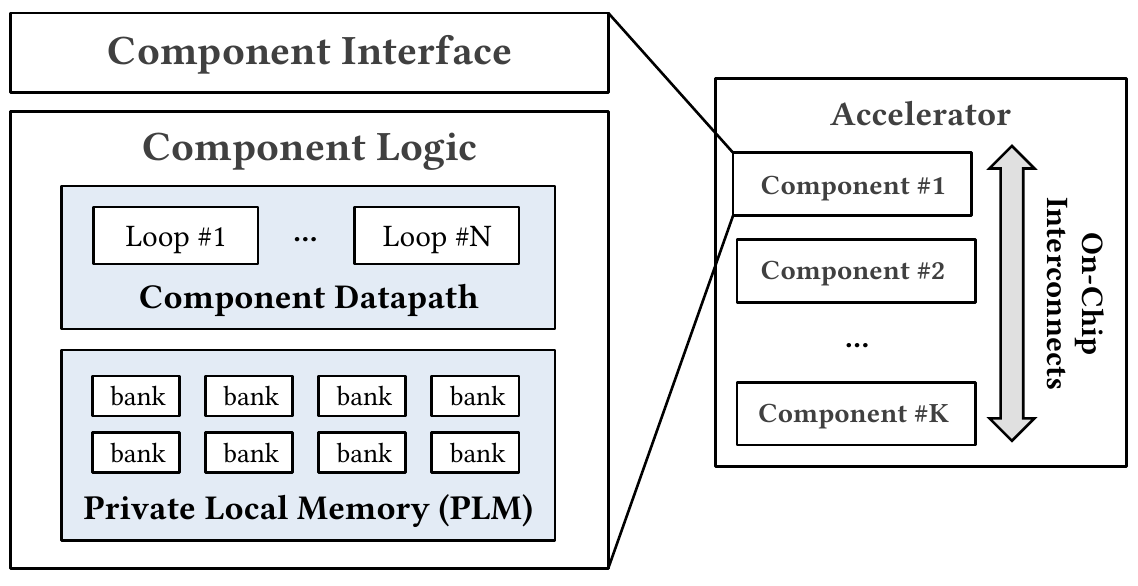}}
\caption{Architecture of a loosely-coupled accelerator.}\label{figure:accelerator}
\end{figure}

\subsection{Hardware Accelerators}\label{section:background:acc}

Several accelerator designs have been proposed in the literature to realize
hardware implementations that execute important computational kernels
more efficiently than corresponding software executions~\cite{
chen_14, zhang_15, liu_fpga2016, reagen2016, kim_tnn17, chen_17}. The accelerators can be located
either inside ({tightly-coupled}) or outside ({loosely-coupled}) the processing
cores~\cite{cota_dac15}. The former class of accelerators is more suitable for
fine-grain computations on small data sets, while the latter is better for
coarse-grain computations on large data sets. We focus on loosely-coupled
accelerators in this paper because the complexity of their design requires a compositional
approach. WAMI is representative  of a set of
classes of applications that can be benefit from the adoption of the
loosely-coupled accelerator model and a compositional design approach.

\subsubsection*{Architecture}

{
We design our accelerators in SystemC.  \figurename~\ref{figure:accelerator}
illustrates their typical architecture.  {They are made of multiple components that
are designed individually to cope with the current limitations of HLS tools in
optimizing complex components. Partitioning the accelerators into multiple
components allows HLS tools to handle them separately, thus reducing the
synthesis time and improving the quality of results.  Each component is
specified as a separated SystemC module and represents a computational block
within the accelerator. The components communicate by exchanging the data
through an on-chip interconnect network that implements transaction-level
modeling (TLM)~\cite{ghenassia2006} channels.  These channels synchronize the
components by absorbing the potential differences in their computational
latencies with a latency-insensitive communication
protocol~\cite{carloni_pieee15}. This ensures that the components of an
accelerator can always be replaced with different Pareto-optimal
implementations without affecting the correctness of the accelerator
implementation. \toolname employs channels with a fixed bitwidth (256 bits) and
does not explore different design alternatives to implement the communication
among the components. It can be extended, however, to support this type of DSE
by using, for example, the \emph{\textbf{X}Knobs}~\cite{piccolboni_hpec17} or
buffer-restructuring techniques~\cite{cong_dac17}.  Each component includes a datapath, which is
organized in a set of loops, to read and store input and output data and to
compute the required functionality.} There are also \emph{private local
memories} (PLMs), or \emph{scratchpads}, where data resides during the
computation. PLMs are multi-bank memory architectures that provide multiple
read and write ports to allow accelerators to perform parallel accesses.  We
generate optimized memories for our accelerators by using the \textsc{Mnemosyne}
memory generator~\cite{pilato_tcad17}.  Several analyses highlight
the importance \linebreak of the PLMs in sustaining the parallel datapath of
accelerators~\cite{Lyons2012, cota_dac15}. PLMs play a key role on the
performance of accelerators~\cite{li_2011}, and they occupy from 40\% to
90\% of the entire area of the components of a given accelerator~\cite{Lyons2012}.
\fillparagraph
}

\begin{figure}
\centering\resizebox{0.65\linewidth}{!}{
\includegraphics{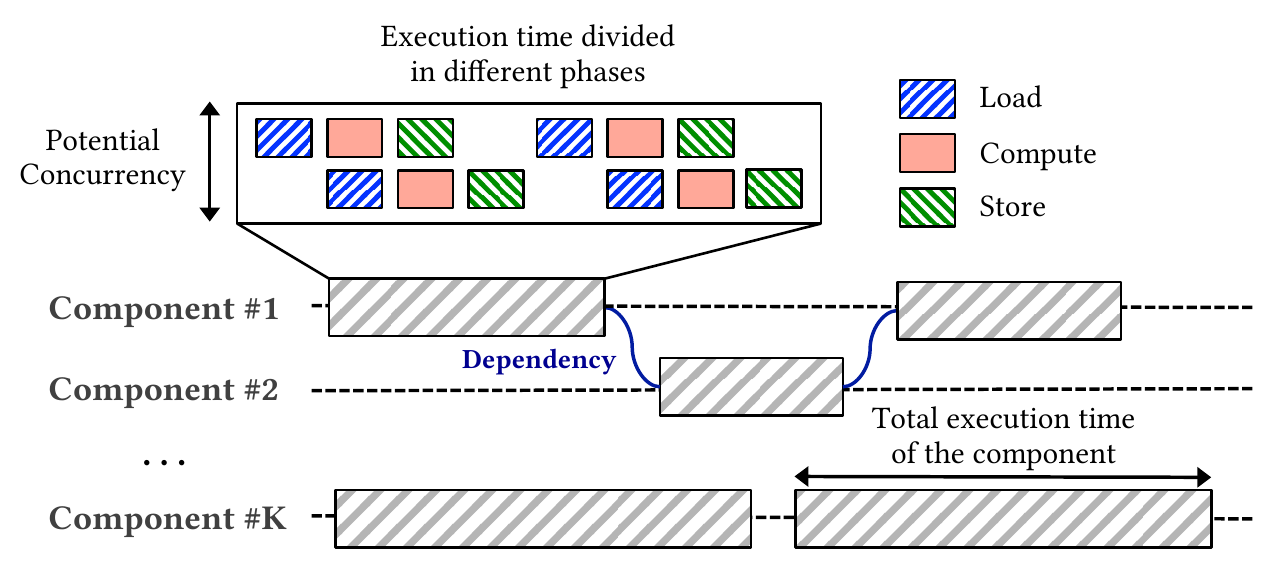}}
\caption{Execution of a loosely-coupled accelerator.}\label{figure:accelerator:exec}
\end{figure}

\subsubsection*{Execution}

{
\figurename~\ref{figure:accelerator:exec} reports an example of execution of an
accelerator made of multiple components. The execution of each component of the
accelerator is divided in three phases (showed on the top of the figure for
Component \#1). In the \emph{load} phase the components communicate with the
on-chip interconnect network to read the input data and store it in the PLMs.
In the \emph{compute} phase the components execute the given functions on 
the data currently stored in the PLMs.
In the \emph{store} phase the components
communicate with the on-chip interconnect network to store the output data
available in the PLMs. These three phases can be pipelined by using techniques
such as ping-pong or circular buffers~\cite{cota_dac15}, as shown on the top of
the figure. After having identified the minimum block of data that is
sufficient to realize the required function in each component, e.g., a
frame, the execution of the components can be: (i) completely overlapped when
there are no dependencies (e.g., Component \#1 and \#K), or (ii) serialized
when a component needs input data from another component to start its
computation (e.g., Component \#1 and \#2). 
\fillparagraph
}
\subsection{Computational Model}\label{section:background:tmg}

{
To formally model the loosely-coupled accelerators we use timed marked 
graphs (TMGs), a subclass of Petri nets (PNs)~\cite{murata_89}. TMGs are commonly
used to perform compositional performance analysis of discrete-event
systems~\cite{campos92}. While TMGs do not allow to capture data-dependent
behaviors, they are a practical model to analyze stream processing accelerators
for many classes of applications, e.g., image and signal processing
applications.
A \emph{PN} is a bipartite graph defined as a tuple $(P,T,F,w,M_0)$,
where $P$ is a set of $m$ places, $T$ is a set of $n$ transitions, $F:(P \times
T) \cup (T \times P)$ is a set of arcs, $w:F \to \mathbb{N}^+$ is an arc
weighting function, and $M_0 \in \mathbb{N}^{m}$ is the initial marking, i.e.
the number of tokens at each $p \in P$. 
{A {PN} is \emph{strongly-connected} if for every pairs of places $p_i$ and
$p_j$ there exists a sequence of transitions and places such that $p_i$ and
$p_j$ are mutually reachable in the net. A {PN} can be organized in a set of
\emph{strongly-connect components}, i.e., the maximal sets of places that are
strongly-connected.}
A \emph{TMG} is a PN such that (i) each place has exactly one input and one
output transition, and (ii) $w:F \to 1$, i.e., {every arc has a weight equal
to $1$}. To measure performance, TMGs are extended with a transition
firing-delay vector $\tau \in \mathbb{R}^{n}$, which represents the
duration of each particular firing. \unskip\parfillskip 0pt \par }
%
%

The \emph{minimum cycle time} of a strongly-connected TMG is defined as:
%
${max}\ \{ D_k/N_k \mid k \in K \}$,
%
where $K$ is the set of cycles of the TMG, $D_k$ is the sum of
the transition firing delays in cycle $k$, and $N_k$ is the number of tokens in cycle
$k$~\cite{ramamoorthy_80}. 
In this paper, we use the TMG model to formally
describe the accelerators. We use the term \emph{system} to indicate a complex
accelerator that is made of multiple \emph{components}. Each component of
the system is represented with a transition in the TMG whose firing delay is
equal to its effective latency. The \emph{effective latency} $\lambda$ of a
component is defined as the product of its clock cycle count and target clock period.
The maximum sustainable {\em effective throughput} $\theta$ of the system is
then the reciprocal of the minimum cycle time of its TMG, if the TMG is
strongly connected. Otherwise, it is the minimum $\theta$ among its
strongly-connected components. We use $\lambda$ and $\theta$ as
performance figures for the single components and the system, respectively. We
use the {\em area} $\alpha$ as the cost metric for both the components and the system.


\section{Motivational Examples}\label{section:example}

Performing an accurate and as exhaustive as possible DSE for a complex hardware
accelerator is a difficult task for three main reasons: 
(i) HLS tools do not always support PLM generation and optimization
{(Section~\ref{section:example:plm}), 
(ii) HLS tools are based on heuristics that make it difficult to configure the
knobs (Section~\ref{section:example:unpred}), 
and (iii) HLS tools do not handle the simultaneous optimization of multiple
components (Section~\ref{section:example:comp}). Next, we detail these issues
with some examples.

%
%

%
%

%
\subsection{Memories}\label{section:example:plm}

The joint optimization of the accelerator datapath and PLM architecture 
is critical for an effective DSE. \figurename~\ref{figure:gradient} depicts the design space of
{\accsmall{Gradient}}, a component we designed for WAMI. The graph reports
different design points, each characterized in terms of area (\textit{mm}$^2$)
and {effective latency (\textit{milliseconds})}, synthesized for an industrial
32nm ASIC technology library. The points with the same color (shape) are
obtained by partially unrolling the loops for different numbers of iterations.
The different colors (shapes) indicate different numbers of ports for the
PLM\footnote{{Here and in the rest of the paper, the number of ports indicates
the number of read ports to the memories containing the input data of the
component and the number of write ports containing the output data of the
component, i.e., the ports that allow parallelism in the compute phase of the
component.}}. By increasing the number of ports, we notice a significant impact
on both latency and area. In fact, multiple ports allow the component to read
and write more data in the same clock cycle, thus increasing the hardware
parallelism. Multi-port memories, however, require much more area since more
banks may be used depending on the given memory-access pattern.
Note that ignoring the role of the PLM limits considerably the design space. By
changing the number of ports of the PLM, we obtain a latency span of
$7.9\times$ and an area span of $3.7\times$.  By using standard dual-port
memories, we have only a latency span of $1.4\times$ and an area span of
$1.2\times$. This motivates the need of considering the
optimization of PLMs in the DSE process.
{\toolname takes into consideration the PLMs by generating optimized memories with 
\textsc{Mnemosyne}~\cite{pilato_tcad17}}.

\begin{figure}[t] 
\centering\resizebox{0.73\linewidth}{!}{
\includegraphics{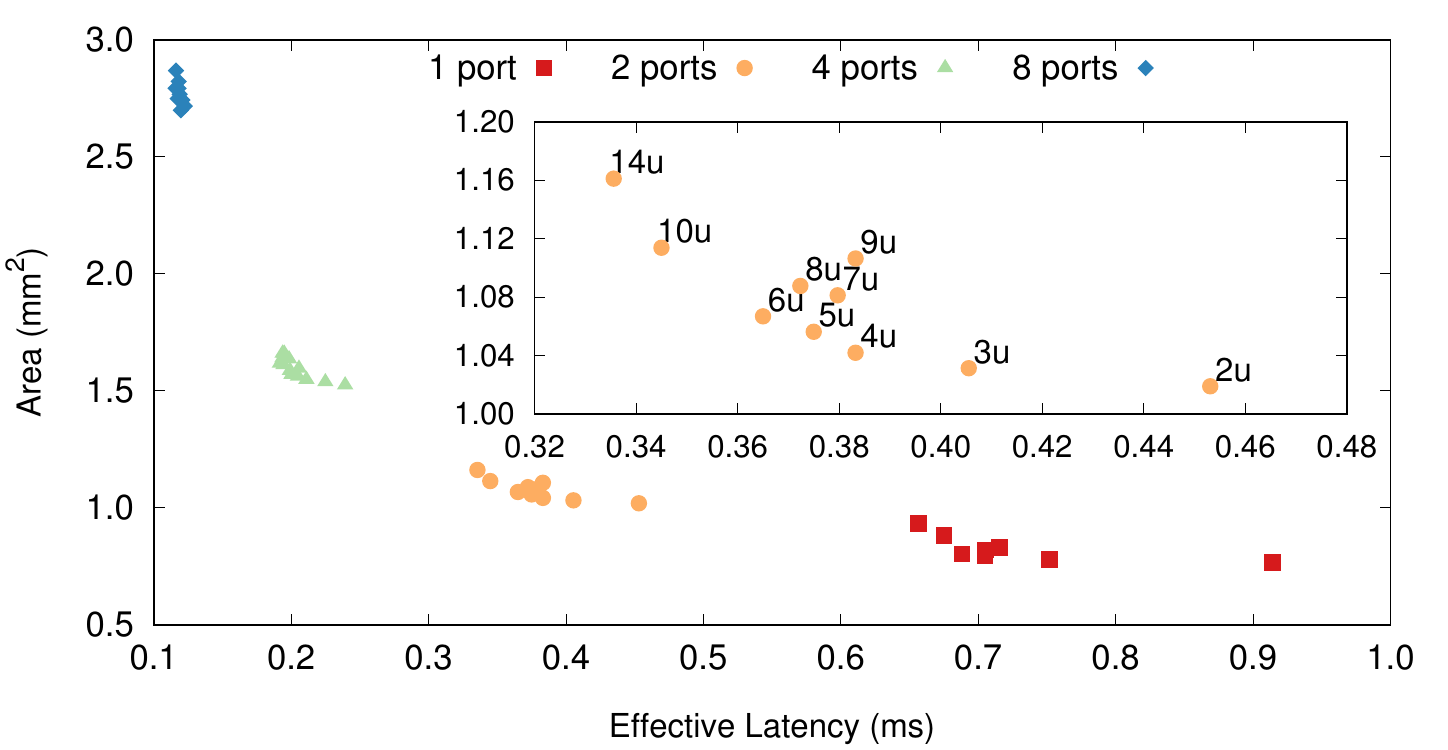}} 
\caption{Example of application of two HLS knobs (number of ports, number of 
 unrolls) to \accsmall{Gradient}, a component of WAMI. The nested graph magnifies the design
points with 2 read and 2 write ports. The numbers indicate the 
numbers of iterations unrolled.}
\label{figure:gradient}
\end{figure}

\subsection{HLS Unpredictability}\label{section:example:unpred}

Dealing with the unpredictability of the HLS tool outcomes is necessary to
remain in the Pareto-optimal regions of the design space~\cite{panda07}. This
is highlighted by the magnified graph in~\figurename~\ref{figure:gradient} that
reports the number of iterations unrolled for each design point of
\accsmall{Gradient}. By increasing the number of iterations unrolled in a loop
for a particular configuration of the PLM ports we expect to obtain design
points that have more area and less latency. In fact, unrolling a loop
increases the number of hardware resources to allow more parallel operations.
However, an effective parallelization is not always guaranteed. Some
combinations of loop unrolling have a negative effect on both latency and area
due to the applications of HLS heuristics (e.g., points $7$u, $8$u and $9$u in
\figurename~\ref{figure:gradient}). In fact, HLS tools need to insert
additional clock cycles in the body of a loop when (i) operation dependencies
are present or (ii) the area is growing too much with respect to the scheduling
metrics they adopt (HLS tools often perform latency-constrained optimizations 
to minimize the area).  This motivates the need of dealing with the HLS 
unpredictability in the DSE process. {\toolname applies synthesis 
constraints to account for the high variability
and partial unpredictability of the HLS tools}. 

\subsection{Compositionality}\label{section:example:comp}

\begin{figure}[t]
\resizebox{0.8\linewidth}{!}{
\begin{tabular}{cc}
\imagetop{\includegraphics{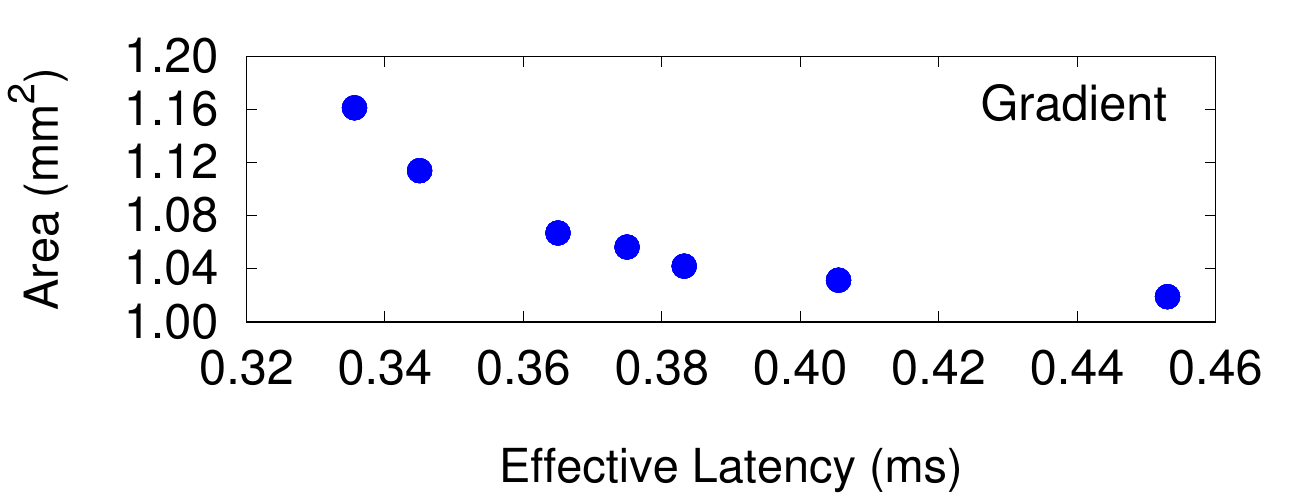}} &
\imagetop{\includegraphics{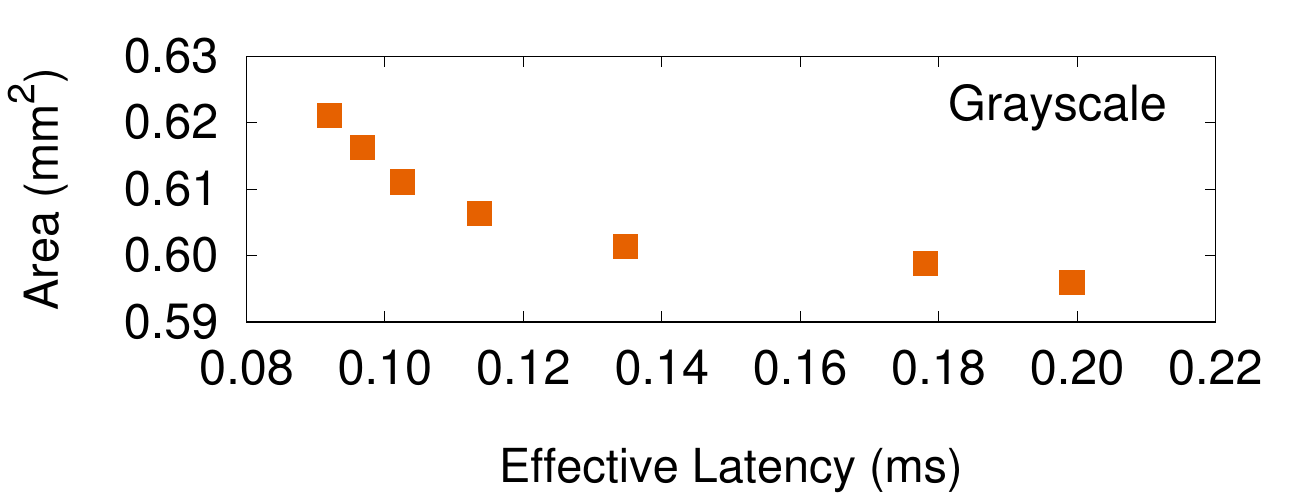}} \\
\end{tabular}}
\resizebox{0.76\linewidth}{!}{
\includegraphics{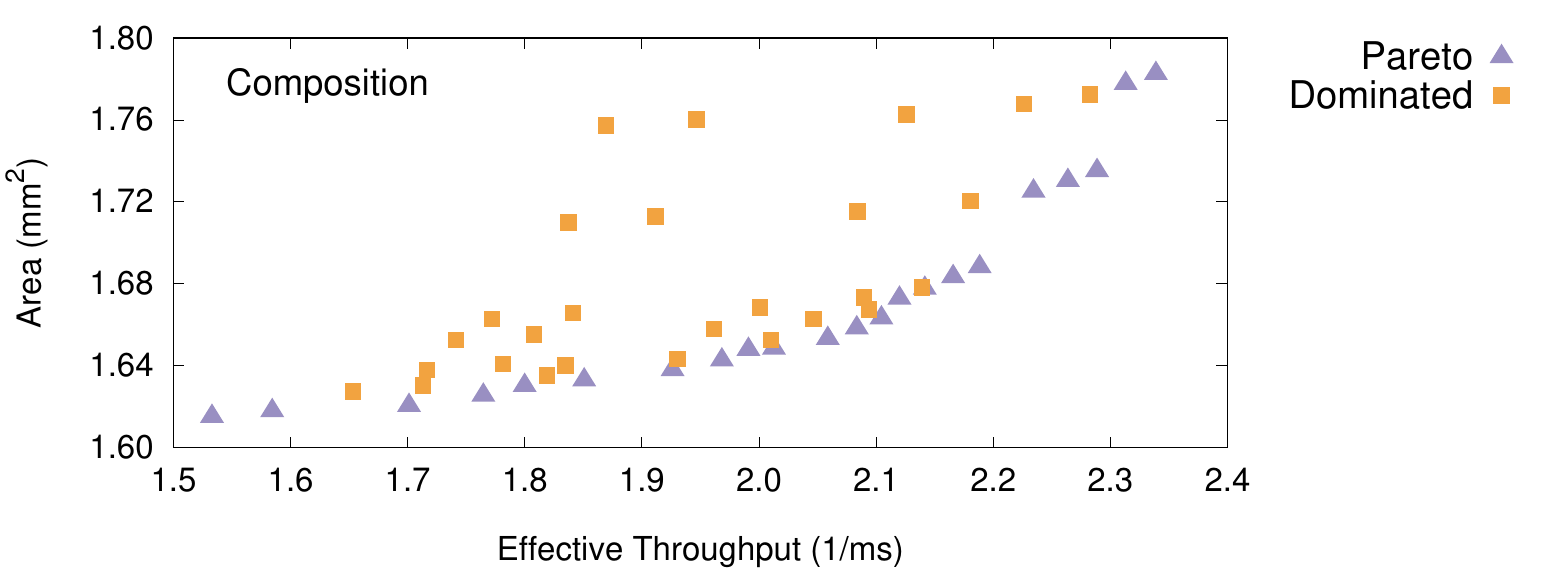}} \\
\caption{Example of composition for \accsmall{Gradient} and
\accsmall{Grayscale}, two components of WAMI. The graphs on the top report some
Pareto-optimal points for the two components. The graph on the bottom shows all
the possible combinations of these components, assuming they are executed
sequentially in a loop.  In the graph of the composition, the effective
throughput is used as the performance metric.}
\label{figure:compositionality}
\end{figure}

Complex accelerators need to be partitioned into multiple components to be
efficiently synthesized by current HLS tools. {This reduces the synthesis time
and improves the quality of results, but significantly increases the DSE effort.}
\figurename~\ref{figure:compositionality} reports a simple example to illustrate
this problem. On the top, the figure reports two graphs representing a small
subset of Pareto-optimal points for \accsmall{Gradient} and
\accsmall{Grayscale}, two components of WAMI. Assuming that they are executed
sequentially in a loop, their aggregate throughput is the reciprocal of the sum
of their latencies. On the bottom, the figure reports all the possible
combinations of the design points of the two components, differentiating the
Pareto-optimal combinations from the Pareto-dominated combinations. These
design points are characterized in terms of area (\textit{mm}$^2$) and
effective throughput (\textit{1/milliseconds}).  In order to find the
Pareto-optimal combinations at the system level, an exhaustive search method
would apply the following steps: (i) synthesize different points for each
component by varying the settings of the knobs, (ii) find the Pareto-optimal
points for each component, and (iii) find the Pareto-optimal combinations of
the components at the system level. This approach is impractical for complex
accelerators. First, step (i) requires to try all the combinations of the knob
settings (e.g., different number of ports and number of unrolls). Second, step
(iii) requires to evaluate an exponential number of combinations at the system
level to find those that are Pareto-optimal. In fact, if we have $n$ components
with $k$ Pareto-optimal points each, then the number of combinations to check
is $\mathcal{O}(k^n)$.  This example motivates the need of a smart
compositional method that identifies the most critical components of an
accelerator and minimizes the invocations to the HLS tool.  In order to do
that, \toolname reduces the number of combinations of knob settings that are
used for synthesis and prioritizes the synthesis of the components depending on
their level of contribution to the effective throughput of the entire
accelerator.


%
%

\section{The COSMOS Methodology}\label{section:design}

As shown in \figurename~\ref{figure:overview}, \toolname consists of the following steps:
\begin{enumerate}[\ \ (1)]
\itemsep 0.2em

\item \emph{Component Characterization} (Section~\ref{section:compdse}): %
in this step \toolname analyzes each component of the system individually; 
for each component it identifies the boundaries of the regions that include the
Pareto-optimal designs; starting from the HLS-ready implementation of each
component (in SystemC), \toolname applies an algorithm that generates knob and memory
configurations to automatically coordinate the HLS and memory generator
tools; the algorithm takes into account the memories of the accelerators 
and tries to deal with the unpredictability of HLS tools;


%
\item \emph{Design-Space Exploration} (Section~\ref{section:system}): %
in this step \toolname analyzes the design space of the entire system; the
system is modeled with a TMG to find the most critical components for the
system throughput; then, \toolname:

\begin{itemize}
\itemsep 0.1em

\item%
formulates a LP problem instance to identify the latency
requirements of each component that ensure the specified
system throughput and minimize the system cost; {this step
is called \emph{Synthesis Planning} (Section~\ref{section:system:planning});} 

\item%
maps the solutions of the LP problem to the knob-setting
space of each component and runs additional synthesis 
to get the RTL implementations of the components;
{this step is called \emph{Synthesis Mapping} 
(Section~\ref{section:system:mapping}).}

\end{itemize}
\end{enumerate}



\section{Component Characterization}\label{section:compdse}


Algorithm~\ref{algorithm:dsecomp} reports the pseudocode used for the
component characterization. The designer provides the clock period, the maximum
number of ports for the PLMs (mainly constrained by the target technology and
the memory generator) and the maximum number of loop unrolls. In order to keep
the delay of the logic for selecting the memory banks negligible, the number of
ports should be a power of two. Note that this constraint can be partially
relaxed without requiring Euclidean division for the selection
logic~\cite{seznec15}.  The number of unrolls depends on the loop complexity.
Loops with few iterations can be completely unrolled, while more complex loops
can be only partially unrolled. In fact, unrolling loops replicates the
hardware resources, thus making the scheduling more complex for the HLS
tool.  The algorithm identifies regions in the design space of the component. A
region includes design points that have the same number of ports.  They are
bounded by an upper-left ($\lambda_{min}, \alpha_{max}$) and a lower-right
($\lambda_{max}, \alpha_{min}$) point.  These regions represent the design
space of the component that will be used for the DSE at the
system level, as explained in Section~\ref{section:system}.

{
\SetAlgoNoEnd
\LinesNumbered
\RestyleAlgo{ruled}
\footnotesize
\begin{figure}[t]
\begin{minipage}{0.5\linewidth}
\removelatexerror
\begin{algorithm}[H]
\KwIn{$clock$, $max\_ports$, $max\_unrolls$}
\KwOut{set of regions $(\lambda_{max}, \alpha_{min}, \lambda_{min}, \alpha_{max})$}
\algrule
\For{\upshape{ports} = 1 {\bfseries up to} $max\_ports$}
{
    \ShowLn\tcp{Identification of max-$\lambda$ min-$\alpha$ point}
    $(\lambda_{max}, \alpha_{min}) $ = {\bfseries hls\_tool}(ports, ports, $clock$)\;
    \ShowLn\tcp{Identification of min-$\lambda$ max-$\alpha$ point}
    \For{\upshape{unrolls} = $max\_unrolls$ {\bfseries down to} \upshape{ports + 1}}{
        $(\lambda_{min}, \alpha_{max})$ = {\bfseries hls\_tool}(unrolls, ports, $clock$)\;
        \lIf{$\lambda$\_constraint$_{\text{\upshape{ports}}}$(\upshape{unrolls}) \textbf{is} sat}{\textbf{break}}
    }
   \ShowLn\tcp{Generation of the PLM of the component}
   $\alpha_{plm}$ = {\bfseries memory\_generator}(ports)\; 
   $\alpha_{min}$ += $\alpha_{plm}$;\ \ $\alpha_{max}$ += $\alpha_{plm}$\;
   \ShowLn\tcp{Save the region of the design space}
   $save$($ports, unrolls, \lambda_{max}, \alpha_{min}, \lambda_{min}, \alpha_{max}$)\;
}
\algrule
\nonl \textcolor{black!70}{\textbf{tool parameters:}} \textbf{hls\_tool}(unrolls, ports, clock)\; 
\nonl \textcolor{black!70}{\textbf{tool parameters:}} \textbf{memory\_generator}(ports)\; 
\caption{\em Component Characterization}\label{algorithm:dsecomp}
\end{algorithm}
\end{minipage}
\end{figure}
}


Algorithm~\ref{algorithm:dsecomp} starts by identifying the lower-right point
of the region. To identify this design point, it sets the number of unrolls equal 
to the current number of ports (line 3). This ensures that all the ports of the
PLM are exploited and the obtained point is not redundant. In fact, this
point cannot be obtained by using a lower number of ports. On the other hand,
finding the upper-left point is more challenging. A complete unroll 
(which could lead to the point with the minimum latency) is
unfeasible in case of complex loops. Indeed, it is not always guaranteed
that, by increasing the number of unrolls, 
the HLS tool returns an implementation of the component that gives
lower latency in exchange for higher area occupation. To overcome these problems,
Algorithm~\ref{algorithm:dsecomp} introduces a {constraint},
$\lambda-constraint$ for the rest of the paper, that defines the maximum
number of states that the HLS tool can insert in the body of a loop. This helps in
constraining the behavior of the HLS tool to be more deterministic and in
removing some of the Pareto-dominated points. 
Thus, Algorithm~\ref{algorithm:dsecomp} uses the following function to estimate the
number of states that should be sufficient to schedule one iteration of
the loop that includes read and write operations:
\begin{equation}\label{eq:lambda}
h_{ports}(unrolls) = \left \lceil \frac{\gamma_{r} * unrolls}{ports} \right \rceil 
        + \left \lceil \frac{\gamma_{w}}{ports} \right \rceil + \eta
\end{equation}
where $\gamma_r$ is the maximum number of read accesses to the same array per
loop iteration, $\gamma_w$ is the maximum number of write accesses to the same
array per loop iteration and, $\eta$ accounts for the latency required to
perform the operations that do not access the PLM.
These parameters are inferred by traversing the control data
flow graph (CDFG) created by the HLS tool for scheduling the lower-right point.
This function is used as an upper bound of the number of states that the HLS
tool can insert. If this upper bound is not sufficient, then the synthesis
fails and the point is discarded. A synthesis run with a lower
number of unrolls is performed to find another point to be used as the upper-left
extreme (lines 5-7). 

\begin{ex}
\figurename~\ref{figure:constraint} shows an example of using the
$\lambda$-constraint. The loop (reported on the left) contains two read
operations to two distinct arrays, i.e., $\gamma_{r} = 1$, and one write
operation, i.e., $\gamma_{w} = 1$. We assume that all the operations that are
neither read nor write operations can be performed in one clock cycle, i.e.,
$\eta = 1$. The two diagrams (on the right) show the results of the scheduling
by using two ports for the PLM and by unrolling two or three
times the loop, respectively. In the first case (unrolls = 2), the HLS tool can schedule all
the operations in a maximum of  $h_2(2) = 3$ clock cycles. Thus, this point would
be chosen by Algorithm~\ref{algorithm:dsecomp} to be used as upper-left extreme.  In the
second case (unrolls = 3), the HLS tool is not able to complete the schedule within
$h_2(3) = 4$ clock cycles (it needs at least 5 clock cycles). {Thus, this point is discarded.}
\end{ex}

\begin{figure}[t]
\centering
\resizebox{0.7\linewidth}{!}{\includegraphics{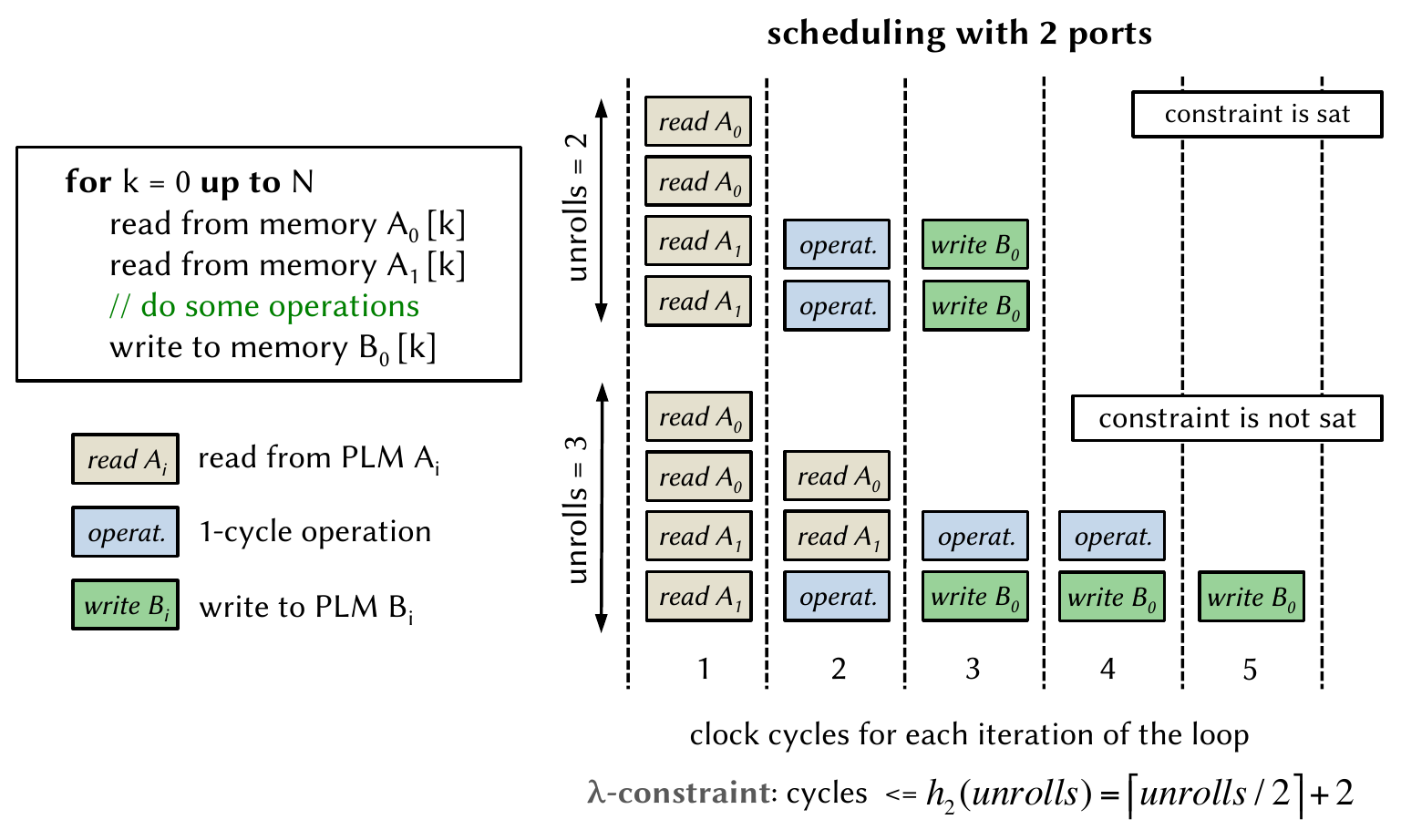}}
\caption{Example of application of the $\lambda$-constraint.}
\label{figure:constraint} 
\end{figure}

Note that the $\lambda$-constraint is not guaranteed to obtain a Pareto-optimal
point due to the intrinsic variability of the HLS results. Still, this point
can serve as an upper bound of the region in the design space. Note also that the
$\lambda-constraint$ cannot be applied to loops that (i) require data from
sub-components through blocking interfaces or (ii) do not present memory
accesses to the PLM.  In these cases, in fact, it is necessary to extend the
definition of the estimation function given in Equation~(\ref{eq:lambda}) to handle such situations.
Alternatively, 
\toolname can optionally run some synthesis in the neighbourhood of the maximum
number of unrolls and use a local Pareto-optimal point as {the upper-left extreme}.

\subsection{Memory Generation}

After the two extreme points of a region have been determined, the algorithm
instructs the memory generator to create the PLM architecture (line 9). 
\toolname uses \textsc{Mnemosyne}~\cite{pilato_tcad17} to generate optimized
PLMs for the components. \textsc{Mnemosyne} has been integrated with the
commercial HLS tool we use for the experimental results
(Section~\ref{section:results}).
The CDFG, created by the HLS tool, is
analyzed to find the arrays specified in the code and their access patterns.
Then, a memory is generated according to these specifications and the area
required for the PLM is added to the logic area reported by the HLS tool (line
10). The memory architecture is tailored to the component needs and is optimized with
respect to the required number of ports and access patterns.  In particular,
given a certain number of ports, \textsc{Mnemosyne} combines several SRAMs,
or BRAMSs in case of FPGA devices, into a multi-bank architecture
(Figure~\ref{figure:accelerator}). Each SRAM (BRAM) provides 2 read/write
ports, thus by combining them in a multi-bank architecture \textsc{Mnemosyne} allows the component to perform multiple
accesses in parallel~\cite{baradaran2008}.  


%



%
%

\section{Design-Space Exploration}\label{section:system}

After the characterization of the single components of a given accelerator,
\toolname uses a LP formulation to find the Pareto-optimal design points at the
system level. The DSE problem at the system level can be formulated as follows:

\begin{problem}\label{problem:dse}
\emph{Given a TMG model of the system where each component has been 
characterized, a HLS tool, and a target granularity $\delta>0$, find 
a \emph{Pareto curve}\ $\alpha$ versus $\theta$ of the system, such that:
\begin{enumerate}[(i)]
\item given two consecutive points $d$, $d'$ on the Pareto curve, they have to
satisfy: $max\ \{d_{\alpha}'/d_{\alpha}-1, d_{\theta}'/d_{\theta}-1\} < \delta$; 
this ensures a maximum distance between two design points on the curve;
\item the HLS tool must be invoked as few times as possible.
\end{enumerate}}
\end{problem}

This formulation is borrowed from~\cite{liu_date12}, where the authors
propose a solution that requires the manual effort of the designers to characterize 
the components. In contrast, \toolname solves this problem by leveraging 
the automatic characterization method in~Section~\ref{section:compdse} 
and by dividing it into two steps: \emph{Synthesis Planning} and \emph{Synthesis Mapping}.

%
%

\subsection{Synthesis Planning}\label{section:system:planning}

Given a strongly-connected system TMG, \toolname uses 
the following $\theta$-constrained cost-minimization LP formulation:
\begin{equation}\label{equation:min_alpha}
\begin{array}{rl}
$min$        & \sum_{i=1}^n f_i(\tau_i) \\
\mbox{s.t.} & A\sigma + M_0/\theta \geq \tau^- \\
            & \tau_{min}^- \leq \tau^- \leq \tau_{max}^-
\end{array}
\end{equation}
where the function $f_i$ returns the implementation cost ($\alpha$)
of the $i$-th component given the firing-delay $\tau_i$ of transition $t_i$,
$\sigma \in \mathbb{R}^n$ is the transition-firing initiation-time vector, 
$M_0\in\mathbb{N}^m$ is the initial marking, $\tau^- \in \mathbb{R}^m$
is the input-transition firing-delay vector, i.e., $\tau_i^-$ is the
firing-delay of the transition $t_k$ entering in place $p_i$ (note that
$\tau_{min}^-$ and $\tau_{max}^-$ correspond to the extreme $\lambda_{min}$
and $\lambda_{max}$ of the components), and $A$ 
is the ${m \times n}$ incidence matrix defined as:
 \begin{equation}
 A[i,j] = \left\{
 \begin{array}{l l}
 +1 & \quad \text{if $t_j$ is an output transition of $p_i$,} \\
 -1 & \quad \text{if $t_j$ is an input transition of $p_i$,} \\
  0 & \quad \text{otherwise.} \\
 \end{array} \right.
 \end{equation}
The objective function minimizes the implementation costs of the components,
while satisfying the system throughput requirements. Given the component extreme latencies
$\lambda_{min}$ and $\lambda_{max}$, it is possible to
determine the values of $\theta_{min}$ and $\theta_{max}$ by labeling the transitions of the
TMG of the system with such latencies. By iterating from $\theta_{min}$ to
$\theta_{max}$ with a ratio of $(1+\delta)$, we can then find the optimal
values of $\lambda$ for the components that solve Problem~\ref{problem:dse}.
This formulation guarantees that the components that are not critical for the
system throughput are selected to minimize their cost.  The cost
functions $f_i$ in Equation~(\ref{equation:min_alpha}) are unknown a-priori, but
they can be approximated  with convex piecewise-linear functions. This
LP formulation can be solved in polynomial time~\cite{convex}, and it
\mbox{can be extended to the case of non-strongly-connected TMGs.}

%
%

\subsection{Synthesis Mapping}\label{section:system:mapping}

Given the optimal values of $\lambda$ of each component that solve
Problem~\ref{problem:dse}, it is necessary to determine the knob settings that
provide the component implementations meeting such requirements.
In other words, we need an inverse function $\phi$ that {maps} the optimal
solutions $\lambda$ to the corresponding values in the knob-setting space of each component. The
solutions of Equation~(\ref{equation:min_alpha}) can require values of $\lambda$
for a component falling inside a certain region. Since we have only the
component implementations for the extreme points of the region (synthesized
with Algorithm~\ref{algorithm:dsecomp}), we need to find the knob settings that
return also the intermediate points.
%
%
%
%
Given the
$\lambda_{target}$ requirement of a component (from
Equation~(\ref{equation:min_alpha})), \toolname first finds the region (returned by
Algorithm~\ref{algorithm:dsecomp}) in which $\lambda_{target}$ falls, i.e.,
$\lambda_{target} \in [\lambda_{min}, \lambda_{max}]$. Then, since every region 
includes design points that have the same number of ports, \toolname
needs only to estimate the number of unrolls to generate a proper knob setting. To do that,
\toolname uses the following modified version of \emph{Amdahl's
Law}~\cite{amdahl_67}:
%
\begin{equation}\label{equ:amdahl2}
  \frac{\lambda_{target}}{\lambda_{max}} = 
    \frac{1}{(1 - \frac{\mu_{target} - \mu_{min}}{\mu_{max} - \mu_{min}}) + 
         \frac{\mu_{target} - \mu_{min}}{\mu_{max} - \mu_{min}} * \frac{\lambda_{max}}{\lambda_{min}}}
\end{equation}
where $\mu_{target}$ is the estimated number of unrolls, while $\mu_{min}$,
$\mu_{max}$ are the numbers of unrolls which correspond to $\lambda_{max}$ and
$\lambda_{min}$, respectively. The only unknown term in this equation is
$\mu_{target}$, i.e., the number of unrolls that can be used to satisfy the
$\lambda_{target}$ requirement.  Thus, \toolname uses the following mapping
function to map the $\lambda$ requirements to the effective number of unrolls:
\newcommand{\numeratora}{\lambda_{min}\lambda_{max}\mu_{max}} 
\newcommand{\numeratorb}{\lambda_{min}\lambda_{max}\mu_{min}} 
\newcommand{\numeratorc}{\lambda_{target}\lambda_{min}\mu_{max}} 
\newcommand{\numeratord}{\lambda_{target}\lambda_{max}\mu_{min}} 
\newcommand{\denominator}{\lambda_{target}(\lambda_{max}-\lambda_{min})} 
%
%
%
%
\begin{equation}\label{equ:amdahl}
\begin{split}
\mu_{target} &=\ \phi(\lambda_{target}, \lambda_{min}, \lambda_{max}, \mu_{min}, \mu_{max} ) = \\
             &=\ \frac{\splitfrac{(\numeratora + \numeratord)\ -}{(\numeratorb + \numeratorc)}}{\denominator}
\end{split}
\end{equation}
This function is derived from Equation~(\ref{equ:amdahl2}), and thus it models the
law of \emph{diminishing returns}. This provides a good approximation of the
number of unrolls because, typically, the relative gains in latency keep 
decreasing as we increase the number of unrolls (Section~\ref{section:results}).
After generating the knob settings by using the mapping function, \toolname 
runs the corresponding synthesis to get (i) the actual values for
$\lambda$ and $\alpha$ and (ii) the RTL implementation of the component.

\begin{figure}[t] 
\centering\resizebox{0.65\linewidth}{!}{
\includegraphics{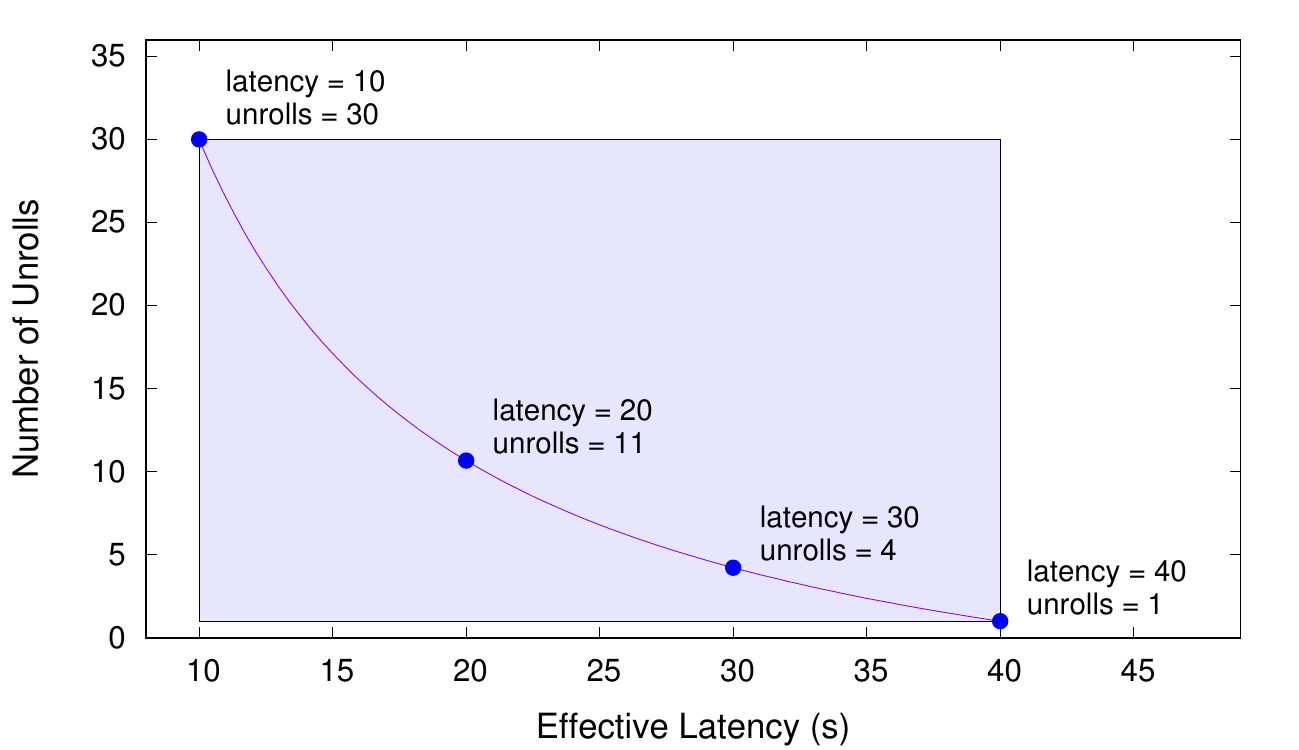}} 
\caption{Example of application of the mapping function~$\phi$.} 
\label{figure:amdahl}
\end{figure}

\begin{ex}\label{example:amdahl}
\figurename~\ref{figure:amdahl} shows an example of a mapping function. 
The lower-right point of the corresponding region has a latency of 40 s, while the
upper-left point has a latency of 10 s, i.e.  $\lambda_{max}$ = 40 s,
$\lambda_{min}$ = \mbox{10 s.} The lower-right point does not unroll the loops,
while the other one unrolls the loops for 30 iterations, i.e., $\mu_{min} = 1$,
$\mu_{max}$ = 30. By using these parameters the graph plots the mapping
function that returns the number of unrolls that should be applied, given a
specific value for the latency ({we apply the ceiling function
to get an integer value}).  For instance, if a point with latency of 20 s
is required, the mapping function returns 11 as the number of unrolls.
Note that by specifying the maximum latency, the function returns the
minimum number of unrolls, while by specifying the minimum latency, it
returns the maximum number of unrolls.
\end{ex}

It is possible that the mapping may fail by choosing a \mbox{value for}
$\mu_{target}$ that does not satisfy the $\lambda$-constraint (Section
\ref{section:compdse}).  In this case, \toolname tries to increase the number of
unrolls to preserve the throughput. Further, if $\lambda_{target}$ is not
included in any region, \toolname uses the slowest point of the next region
that has a larger number of ports.  This does not require a synthesis run ({because that point
has been synthesized during the characterization}), and
it is a conservative solution because, as in the case of failure of the
$\lambda$-constraint, we are willing to \mbox{trade area to preserve the throughput. }


%
%


\section{Experimental Results}\label{section:results}


We implement the \toolname methodology with a set of tools and scripts to
automatize the DSE. {Specifically, \toolname includes: (i) \textsc{Mnemosyne}
\cite{pilato_tcad17} to generate multi-bank  memory architectures as described
in Section~\ref{section:compdse}, (ii) a tool to extract the information
required by \textsc{Mnemosyne} from the database of the HLS tool,} (iii) a
script to run the synthesis and the memory generator according to
Algorithm~\ref{algorithm:dsecomp}, (iv) a program that creates and solves the
LP model by using the GLPK Library\footnote{{ GLPK (GNU Linear Programming
Kit): {\url{https://www.gnu.org/software/glpk/}}}}
(Section~\ref{section:system:planning}), and (v) a tool that maps the LP
solutions to the HLS knobs and runs the synthesis
(Section~\ref{section:system:mapping}).

\begin{figure}
\includegraphics[width=0.6\linewidth]{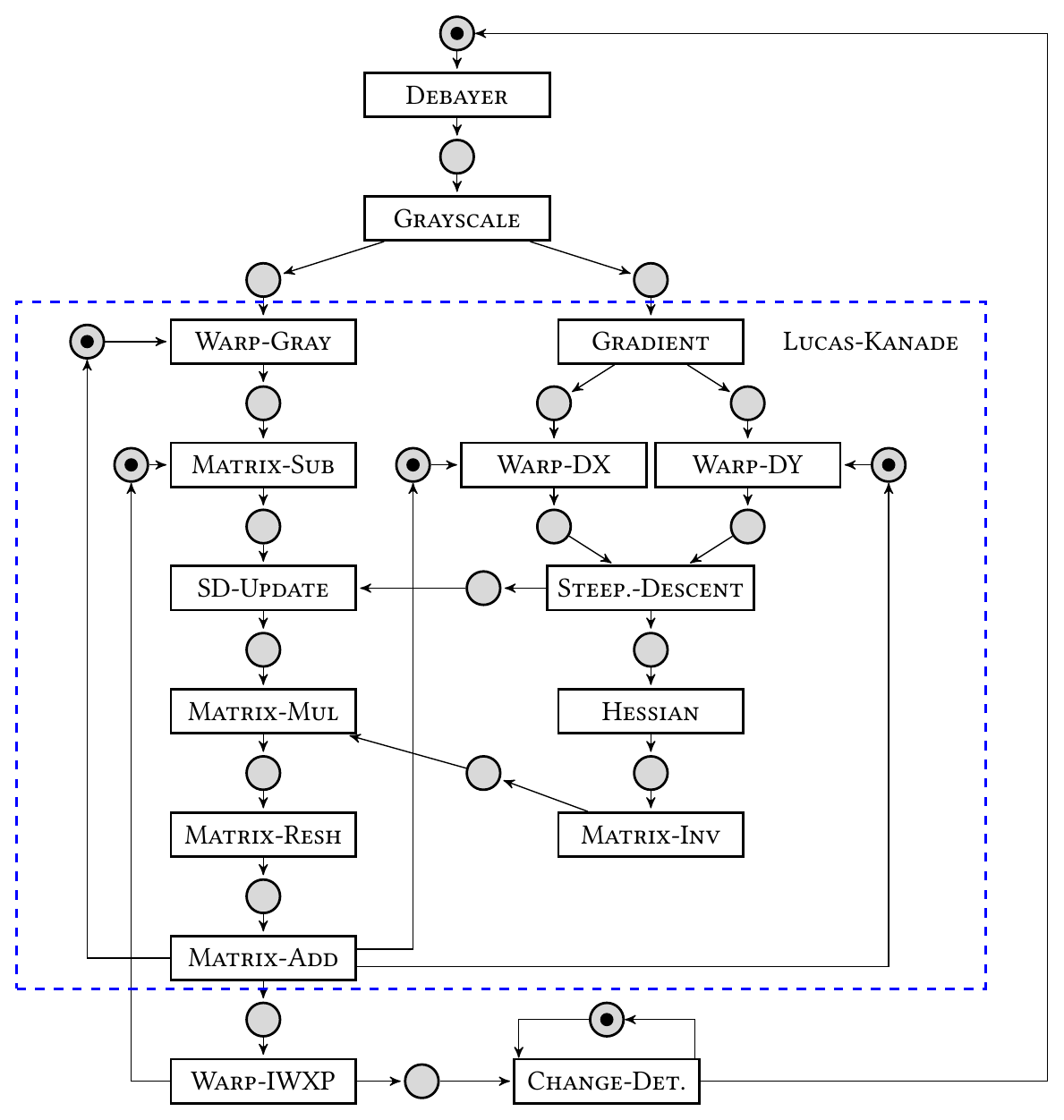}
\caption{TMG modeling the WAMI application.}
\label{figure:wamiapp}
\end{figure}


We evaluate the effectiveness and efficiency of \toolname by considering the
WAMI application~\cite{porter10} as a case study. The original specification of
the WAMI application is available in C in the PERFECT Benchmark
Suite~\cite{perfect-suite-man}. Starting from this specification, we design a
{SystemC} accelerator to be synthesized with a commercial HLS tool, i.e.,
Cadence C-to-Silicon. {We use an industrial $32$nm ASIC technology as target
library\footnote{{ Note that {\scriptsize COSMOS} can be used for FPGA-based
designs as well. It is sufficient to (i) modify the target library used by the
HLS tool and (ii) instructs the memory generator to generate memories by using
the BRAM blocks available in FPGA devices (instead of the SRAM blocks of ASIC
technologies).}}.} We choose the WAMI application as our case study due to (i)
the different types of computational blocks it includes and (ii) its
complexity.  The heterogeneity of its computational blocks allows us to develop
different components for each block and show the vast applicability of
\toolname.  The C specification is roughly $1000$ lines of code. The specification
of our accelerator design is roughly $7000$ lines of SystemC code.

%
%

\subsection{Computational Model}

We model the WAMI application as a loosely-coupled accelerator.
\figurename~\ref{figure:wamiapp} illustrates the resulting TMG model of the
accelerator. The WAMI specification includes four main components: (i)
\accsmall{Debayer} for image filtering, (ii) \accsmall{Grayscale} for
RGB-to-Grayscale color conversion, (iii) \accsmall{Lucas-Kanade} for the image
alignment, and (iv) the \accsmall{Change-Detection} classifier. We partition
\accsmall{Lucas-Kanade} into many sub-components to further increase the
hardware parallelism.  \accsmall{Matrix-Inv} is executed in software to
preserve the floating-point precision. Therefore, it is modeled with a fixed effective
latency during the DSE process.

%
%

\subsection{Component Characterization}

\begin{table}
\centering
\resizebox{0.45\linewidth}{!}{
\begin{tabular}{rccccc}
\toprule%
& 
\multicolumn{3}{c}{\small\bfseries COSMOS}
& 
\multicolumn{2}{c}{\small\bfseries No Memory} \\
\cmidrule(rl){2-4} \cmidrule(rl){5-6}
{\small\bfseries Component} &
{\small $reg$}   &
$\lambda_{span}$ & 
$\alpha_{span}$  &
$\lambda_{span}$ & 
$\alpha_{span}$ \\ \midrule%
\accfoot{Debayer}           & 3 & $2.89\times$ & $1.99\times$ & $1.04\times$ & $1.36\times$ \\
\accfoot{Grayscale}         & 4 & $6.91\times$ & $3.41\times$ & $2.75\times$ & $1.14\times$ \\
\accfoot{Gradient}          & 4 & $7.89\times$ & $3.65\times$ & $1.39\times$ & $1.22\times$ \\
\accfoot{Hessian}           & 4 & $7.70\times$ & $7.30\times$ & $1.44\times$ & $1.30\times$ \\
\accfoot{SD-Update}         & 4 & $9.87\times$ & $2.01\times$ & $2.78\times$ & $1.79\times$ \\
\accfoot{Matrix-Sub}          & 4 & $2.75\times$ & $3.98\times$ & $1.88\times$ & $1.05\times$ \\
\accfoot{Matrix-Add}        & 3 & $1.53\times$ & $1.01\times$ & $1.26\times$ & $1.01\times$ \\
\accfoot{Matrix-Mul}       & 3 & $2.88\times$ & $3.05\times$ & $1.92\times$ & $1.14\times$ \\
\accfoot{Matrix-Resh}    & 1 & $1.02\times$ & $1.04\times$ & $1.02\times$ & $1.04\times$ \\
\accfoot{Steep.-Descent}    & 1 & $1.95\times$ & $1.46\times$ & $1.95\times$ & $1.46\times$ \\
\accfoot{Change-Det.}  & 1 & $2.21\times$ & $1.04\times$ & $2.21\times$ & $1.04\times$ \\
\accfoot{Warp}              & 1 & $1.09\times$ & $1.03\times$ & $1.09\times$ & $1.03\times$ \\
\midrule%
{\small\bfseries Average }& - & $4.06\times$ & $2.58\times$ & $1.73\times$ & $1.22\times$  \\
\bottomrule%
\end{tabular}}
\vspace{0.4cm}
\caption{Characterization of the components for WAMI. The table reports the
differences in latency ($\lambda$) and area ($\alpha$) span when memory is
considered ({\small\bfseries COSMOS}) or not ({\small\bfseries  No
Memory}). For \toolname, $reg$ indicates the number of regions found with Algorithm~\ref{algorithm:dsecomp}.}\label{table:characterization}
\end{table}

{
\toolname applies Algorithm~\ref{algorithm:dsecomp}
(Section~\ref{section:compdse}) to characterize the components of the system.
\tablename~\ref{table:characterization} reports the results of the
characterization for the WAMI accelerator: the algorithm used by \toolname
(\textbf{\small COSMOS}) is compared with the case in which memory is not
considered in the characterization (\textbf{\small No Memory}). In the latter
case, we assume to have only standard dual-port memories. For each component,
the table reports the latency span ($\lambda_{span}$), i.e., the ratio between
the maximum latency and the minimum latency, the area span ($\alpha_{span}$),
i.e., the ratio between the maximum area and the minimum area. For \toolname,
the table shows also the total number of regions identified by the algorithm
($reg$). For Algorithm~\ref{algorithm:dsecomp} we use a number of ports in 
the interval $[1, 16]$ and a maximum number of unrolls in the interval $[8, 32]$, depending on the components.
\toolname guarantees overall a richer DSE, as evidenced by the average results.
For some components the algorithm extracts only one region because multiple
ports can incur in additional area for no latency gains. This happens when (i)
the algorithm cannot exploit multiple accesses in memory, or (ii) the data is
cached into local registers which can be accessed in parallel in the same clock
cycle, e.g., for \accsmall{Change-Detection}. On the other hand, in most cases
\toolname provides significant gains in terms of area and latency spans
compared to a DSE that does not consider the memories.
\fillparagraph
}

\figurename~\ref{figure:characterization} shows the design space of four representative
components of WAMI. The rectangles in the figures are the regions found by
Algorithm~\ref{algorithm:dsecomp}.  For completeness, in addition to the design
points corresponding to the extreme points of the regions, the graphs show also
the intermediate points that could be selected by the mapping function. The
small graphs on the right magnify the corresponding regions reported on the
left. As in the examples discussed in Section~\ref{section:example}, increasing
the number of ports has a significant impact on the DSE, while loop unrolling has a
local effect within each region. Another aspect that is common among many
components is that the regions become smaller as we keep increasing the number
of ports. For example, for \accsmall{Grayscale} in
\figurename~\ref{figure:characterization} (c), we note that by increasing the number
of ports, we reach a point where the gain in latency is not significative.
This effect, called diminishing returns~\cite{amdahl_67}, is the same
effect that can be observed in the parallelization of software algorithms.  In
some cases, changing the ports increases only the area with no latency gains as
discussed in the previous paragraph. This is highlighted in
\figurename~\ref{figure:characterization} (d), where for \accsmall{Change-Detection}
we report two additional regions with respect to those specified
in~\tablename~\ref{table:characterization}.  The diminishing-return effect can
also be observed by increasing the number of unrolls inside a region, e.g.,
\figurename~\ref{figure:characterization} (b). This is why \toolname exploits
Amdahl's Law (Section~\ref{section:system:mapping}).  On the other hand, we
notice some discontinuities of the Pareto-optimal points within some regions,
e.g., the region in the bottom-right corner of \figurename~
\ref{figure:characterization} (a). Even by applying the $\lambda-constraints$
(Section~\ref{section:compdse}) it is not possible to completely discard the
Pareto-dominated implementations. In fact, by further restricting the imposed
constraints, i.e., by reducing the number of states that the HLS tool can
insert in each loop, we observe that also the Pareto-optimal implementations
are discarded. Thus, it is not always possible to obtain a curve composed only
of Pareto-optimal points within a certain region. Finally, the Pareto-optimal
points outside the regions are not discarded by \toolname. They can be chosen
when it is necessary to perform the mapping (Section~\ref{section:system:mapping}).

\begin{figure*}[!hp]
\resizebox{\linewidth}{!}{
\begin{tabular}{c}
\vspace{-0.5cm}%
\subfigure[{\sc\normalsize Debayer}]{
\begin{tabular}{c@{\hspace{0.2cm}}c}
\imagetop{\includegraphics[width=0.56\linewidth]{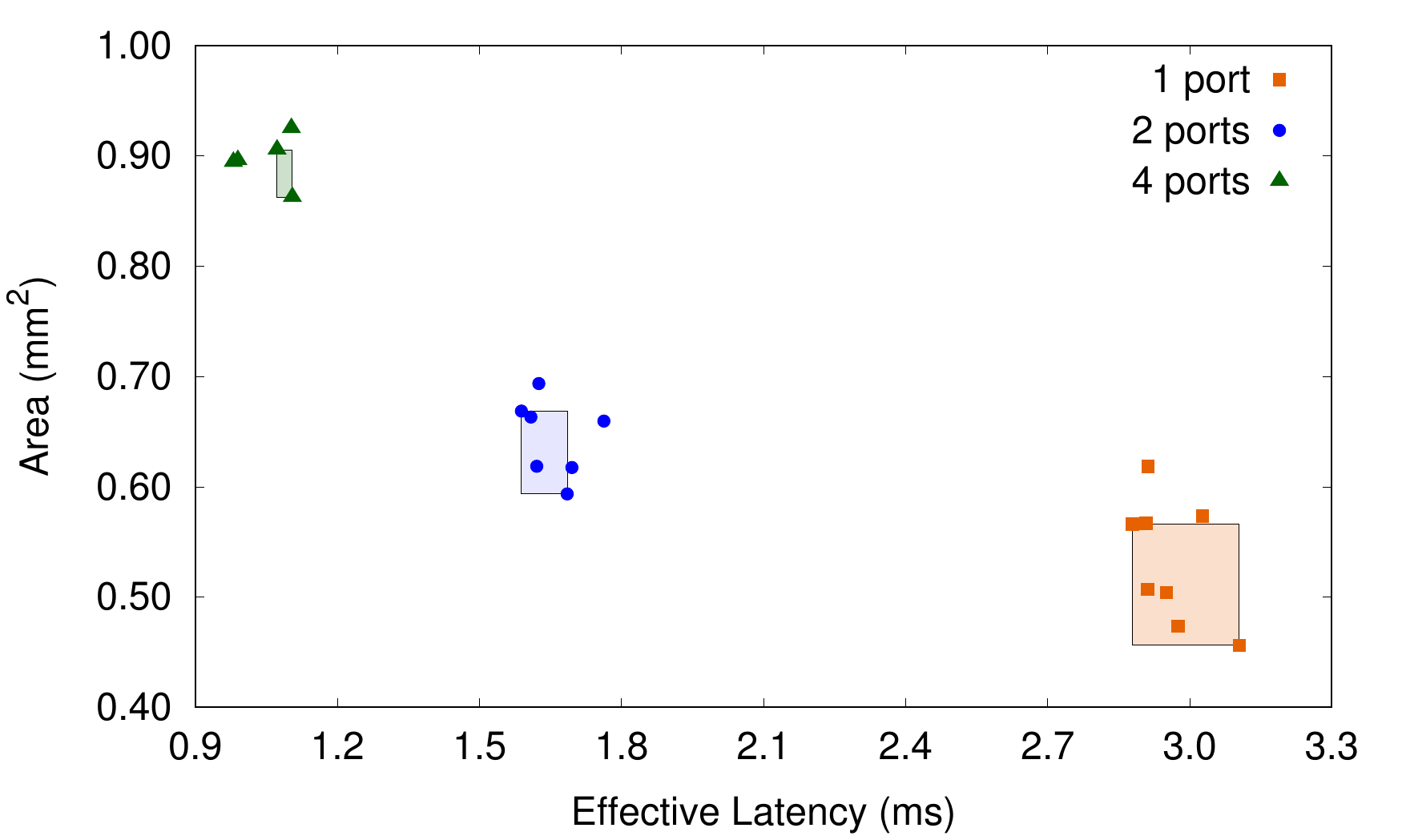}} &
\imagetop{\begin{tabular}{c@{\hspace{0.2cm}}c}
\includegraphics[width=0.25\linewidth]{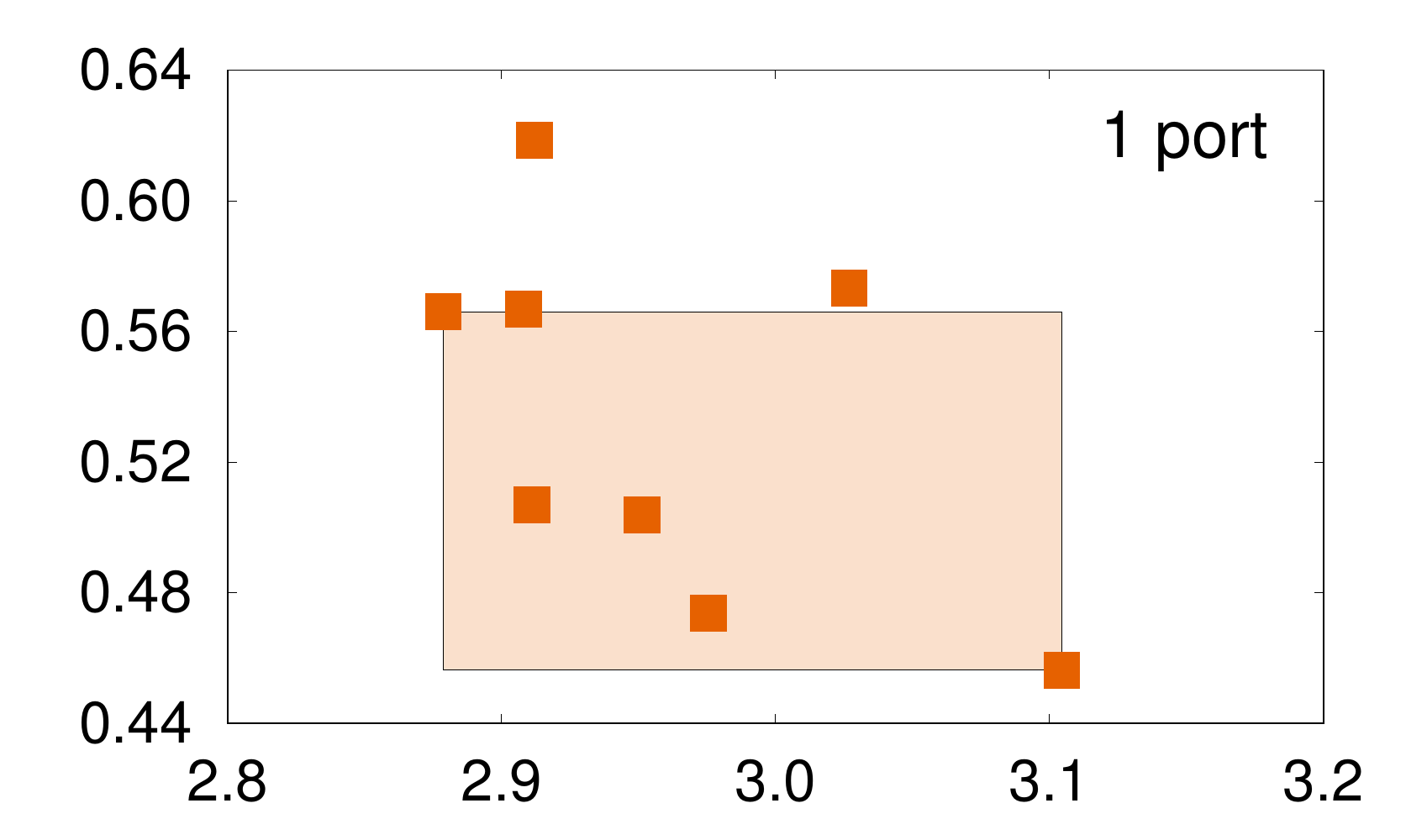} &
\includegraphics[width=0.25\linewidth]{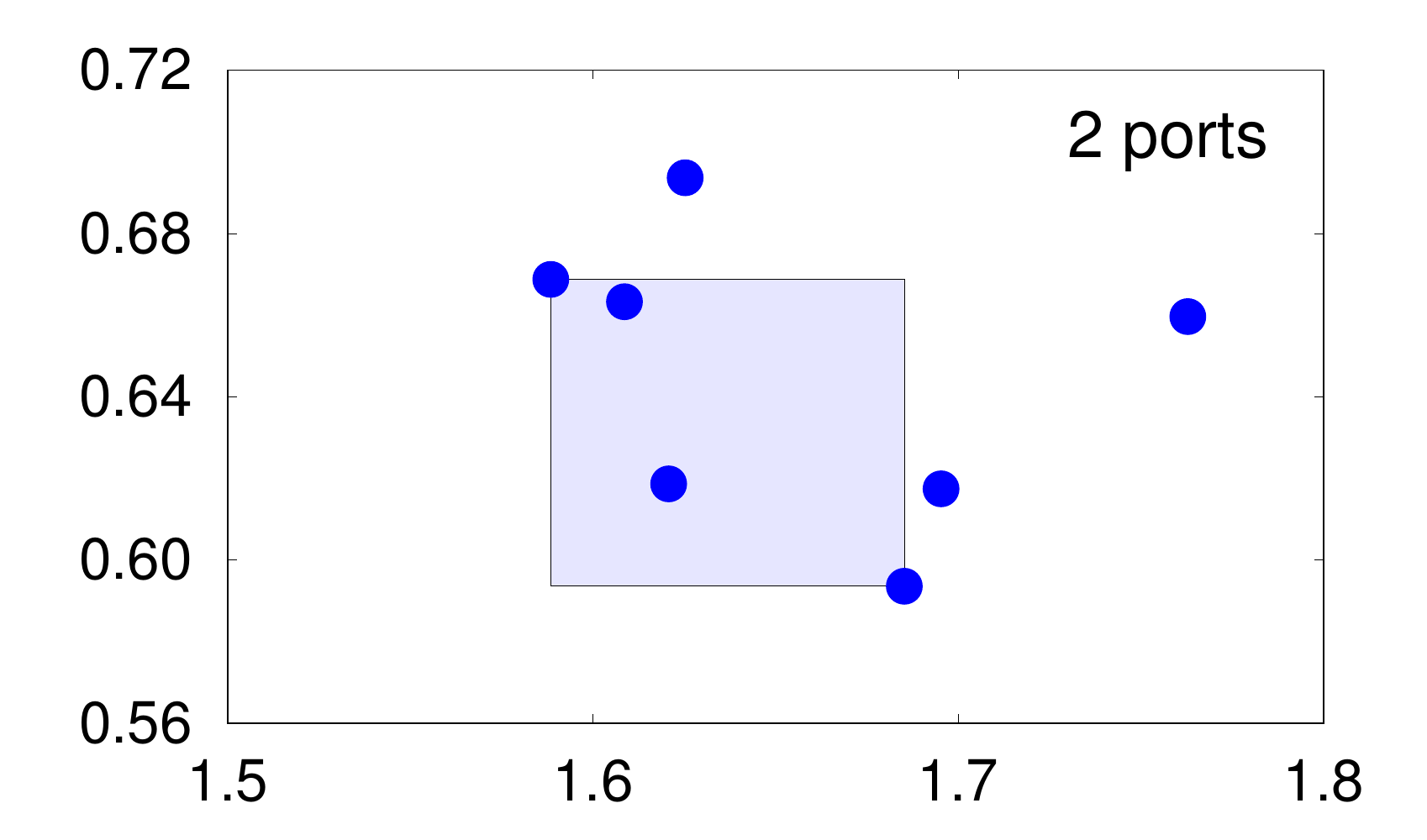} \\
\includegraphics[width=0.25\linewidth]{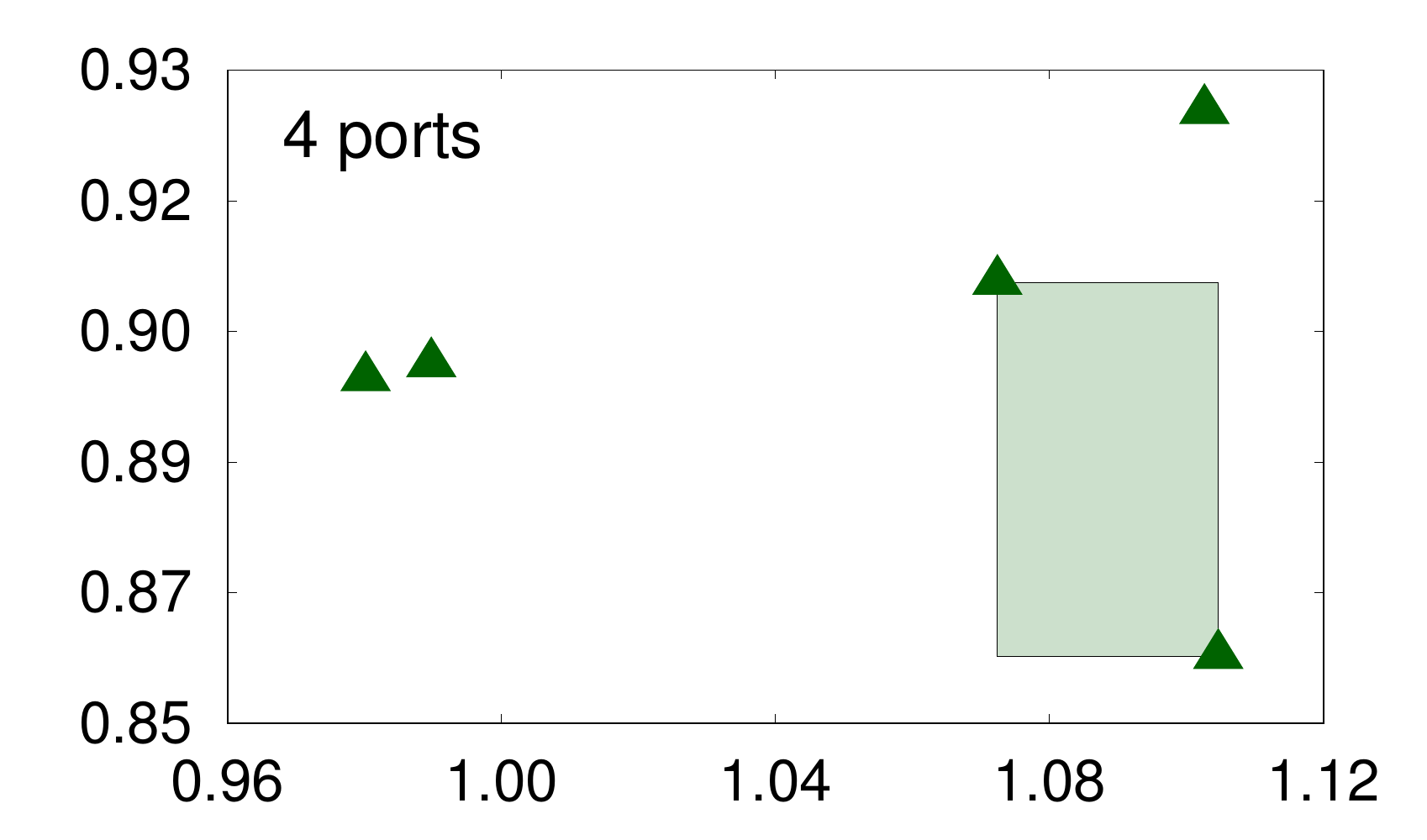} \\
\end{tabular}}
\end{tabular}}\\
\vspace{-0.5cm}%
\subfigure[{\sc\normalsize Hessian}]{
\begin{tabular}{c@{\hspace{0.2cm}}c}
\imagetop{\includegraphics[width=0.56\linewidth]{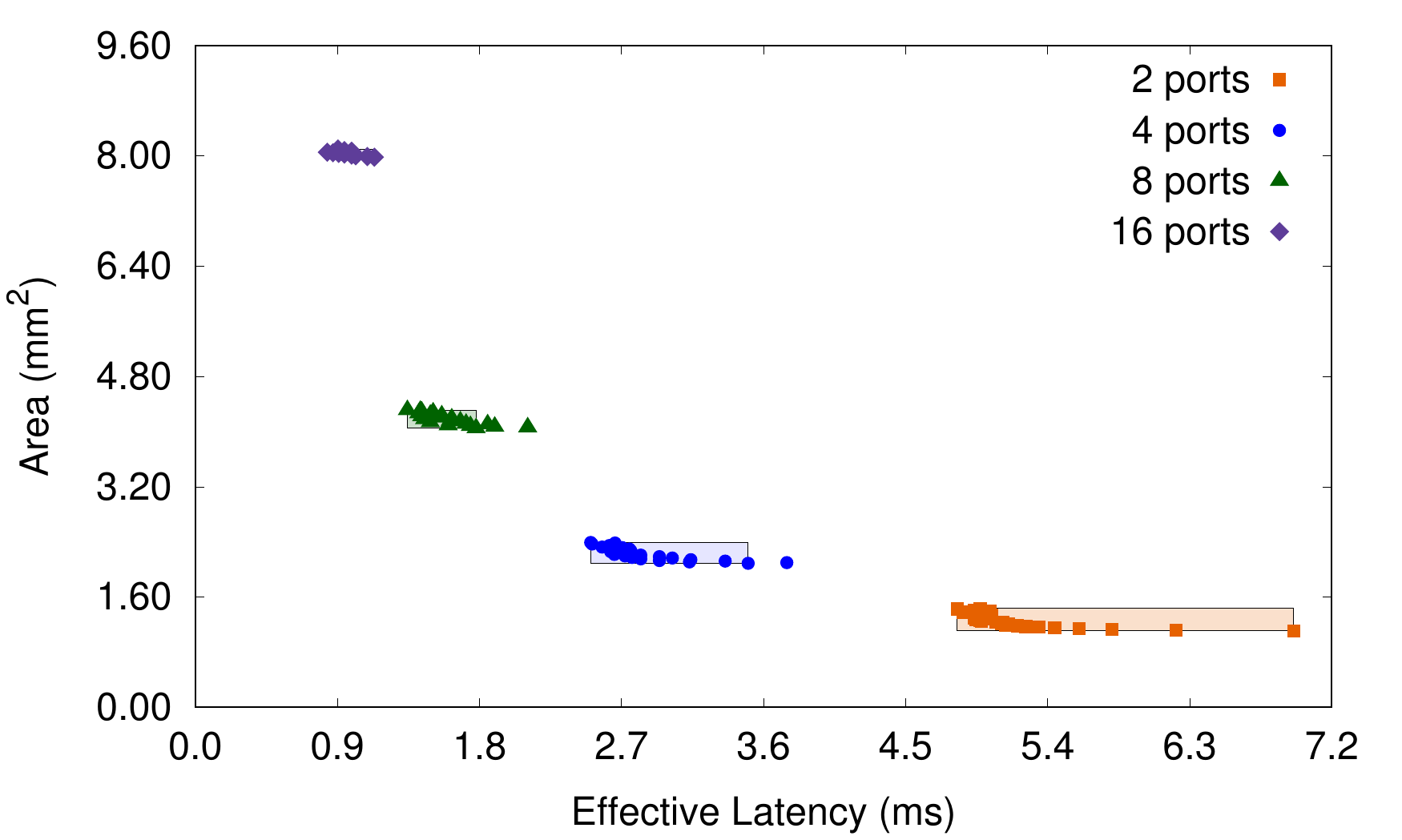}} &
\imagetop{\begin{tabular}{c@{\hspace{0.2cm}}c}
\includegraphics[width=0.25\linewidth]{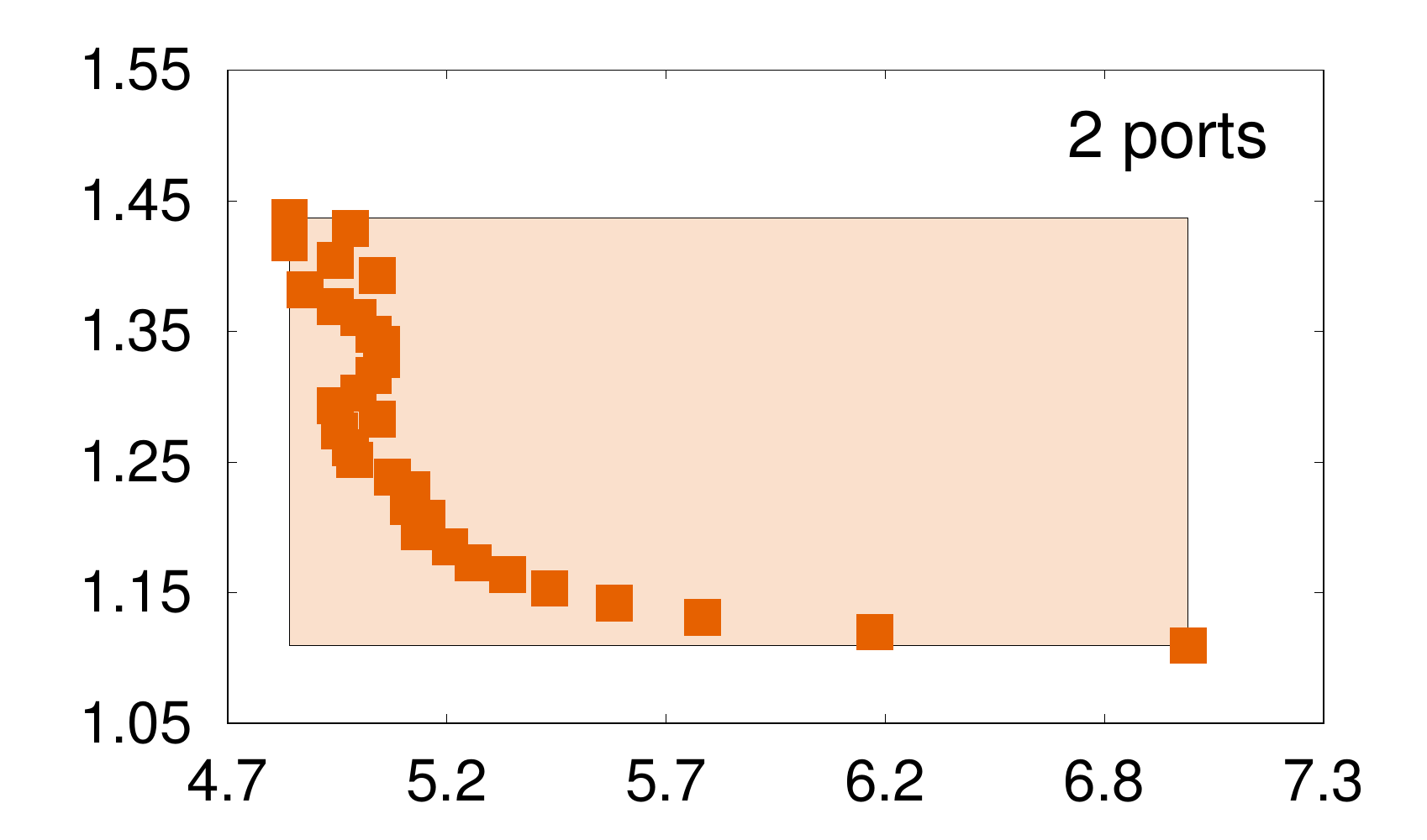} &
\includegraphics[width=0.25\linewidth]{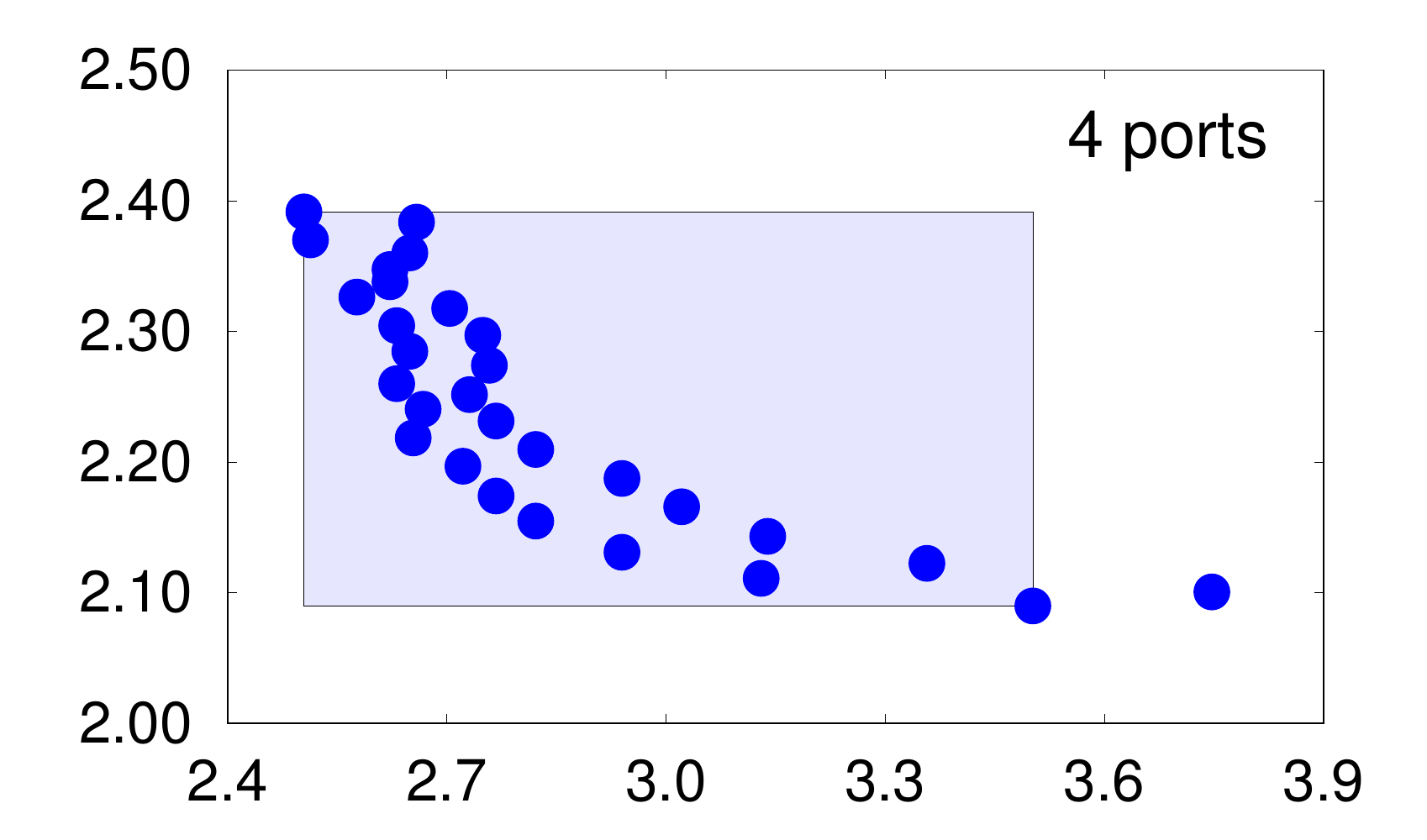} \\
\includegraphics[width=0.25\linewidth]{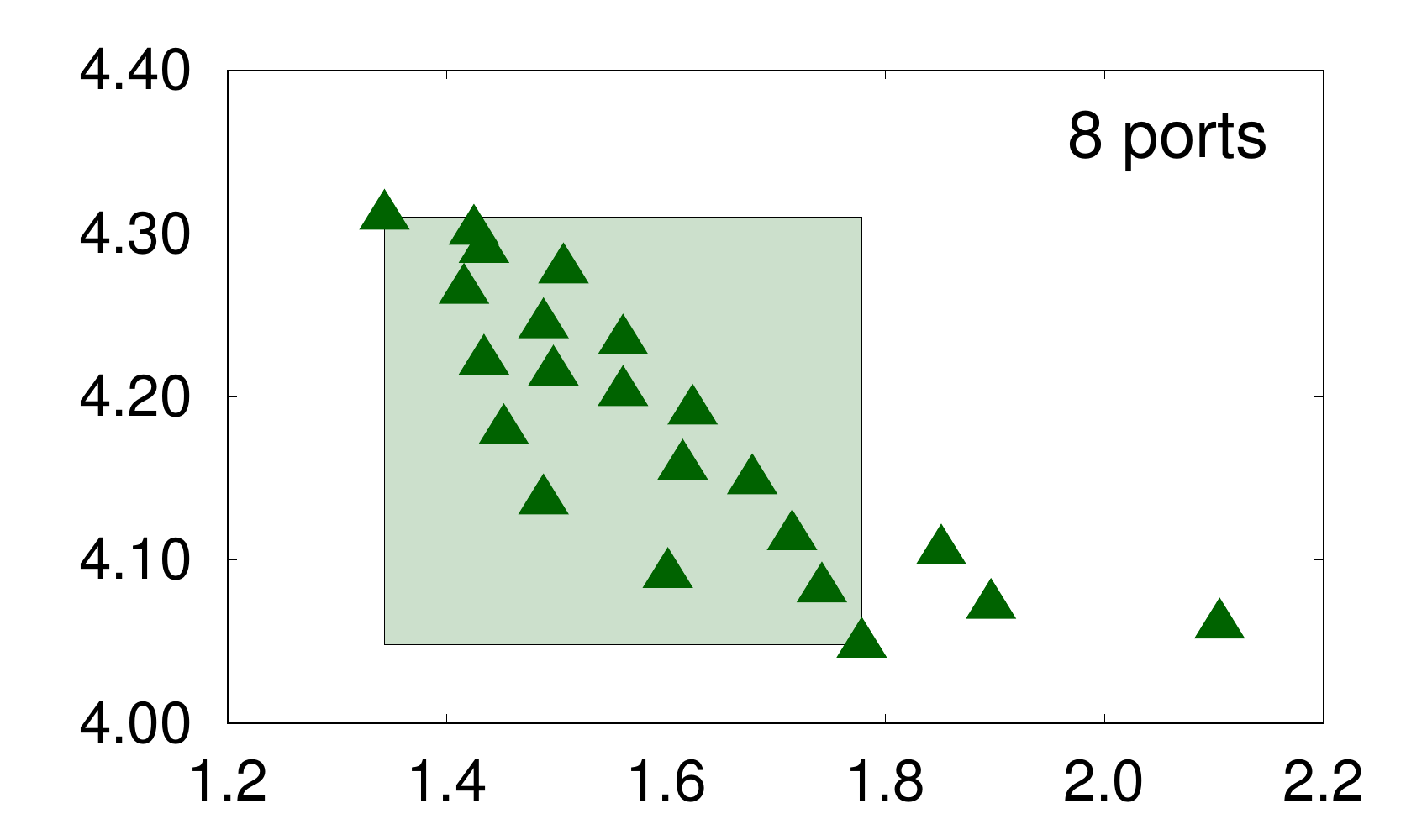} &
\includegraphics[width=0.25\linewidth]{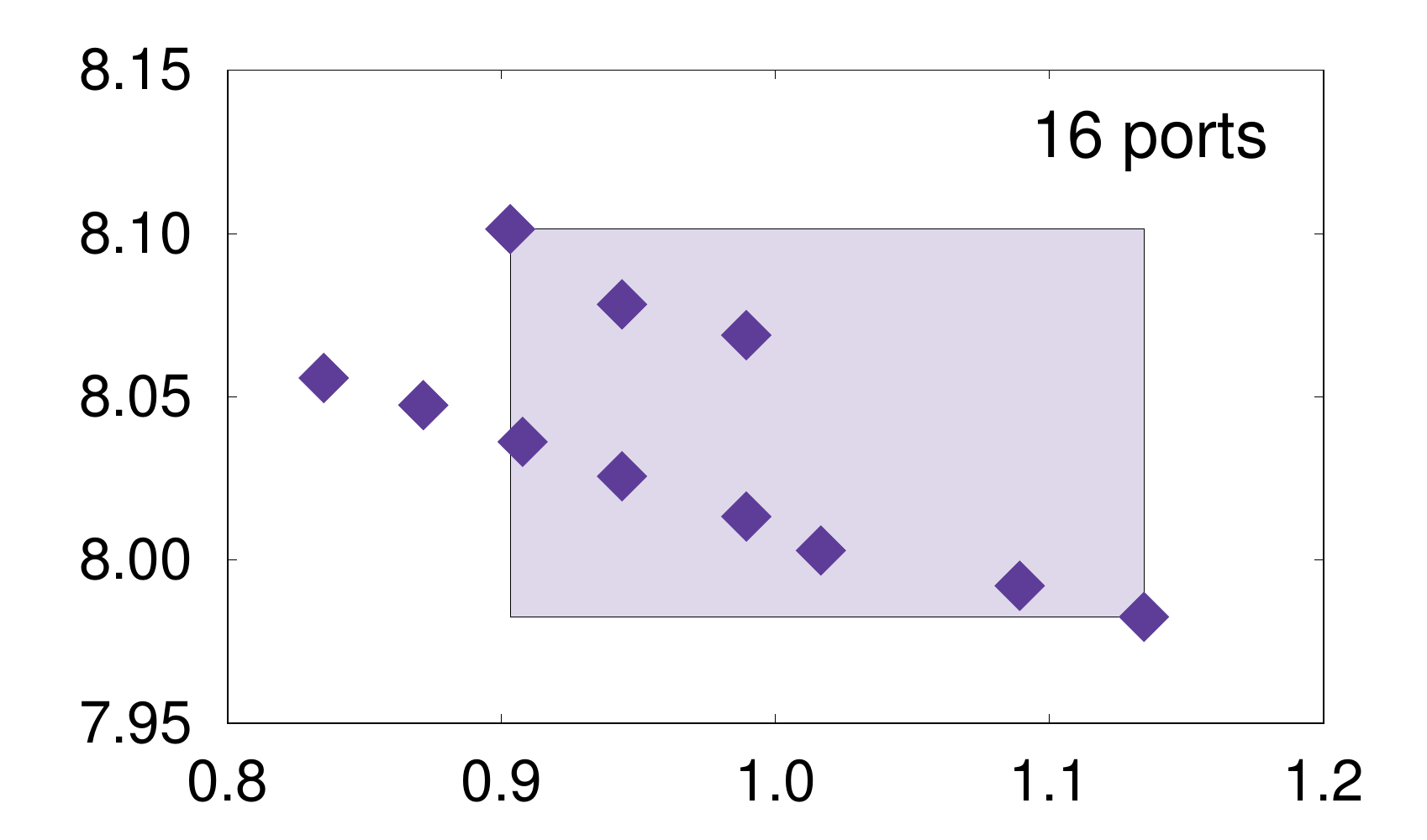} \\
\end{tabular}}
\end{tabular}}\\
\vspace{-0.5cm}%
\subfigure[{\sc\normalsize Grayscale}]{
\begin{tabular}{c@{\hspace{0.2cm}}c}
\imagetop{\includegraphics[width=0.56\linewidth]{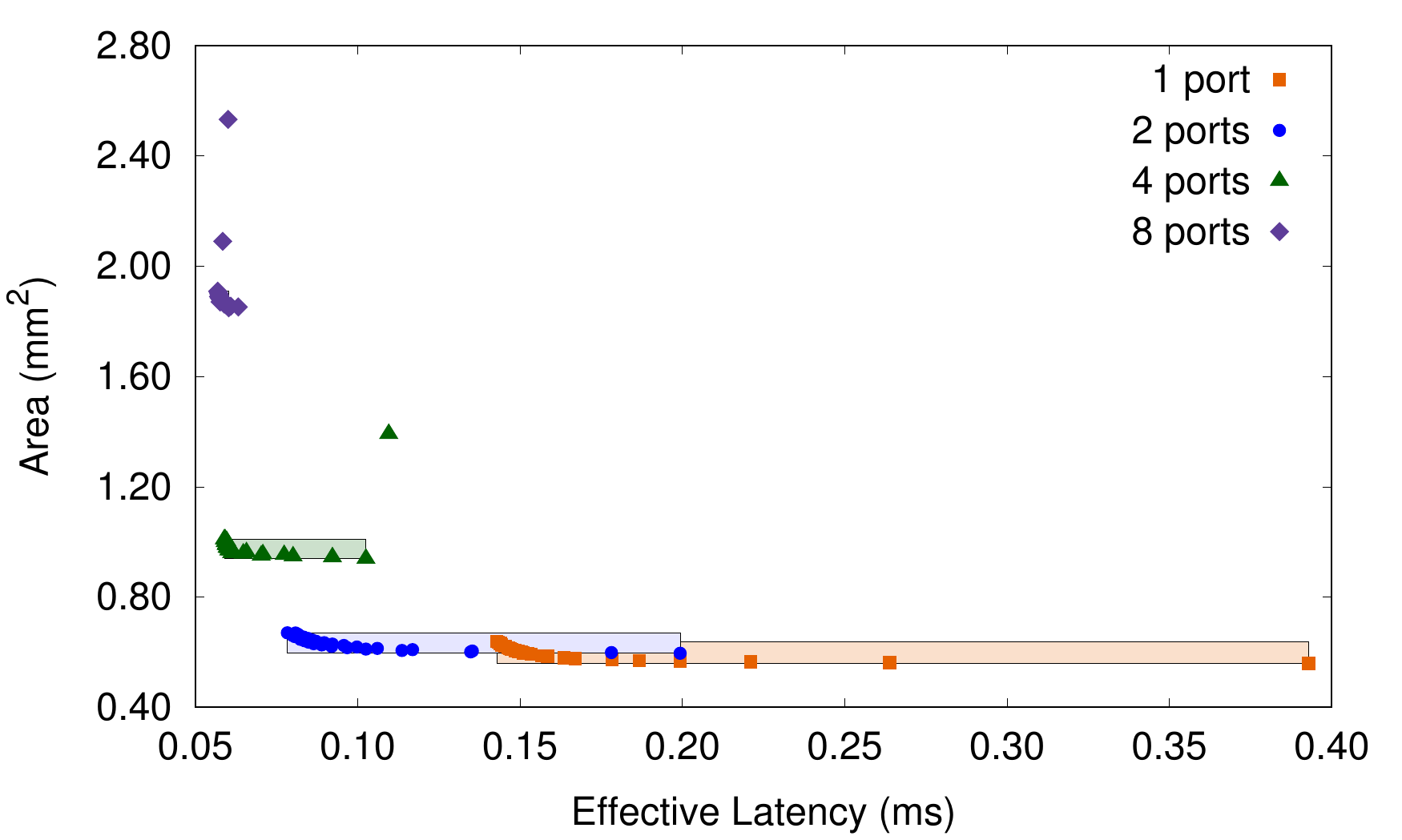}} &
\imagetop{\begin{tabular}{c@{\hspace{0.2cm}}c}
\includegraphics[width=0.25\linewidth]{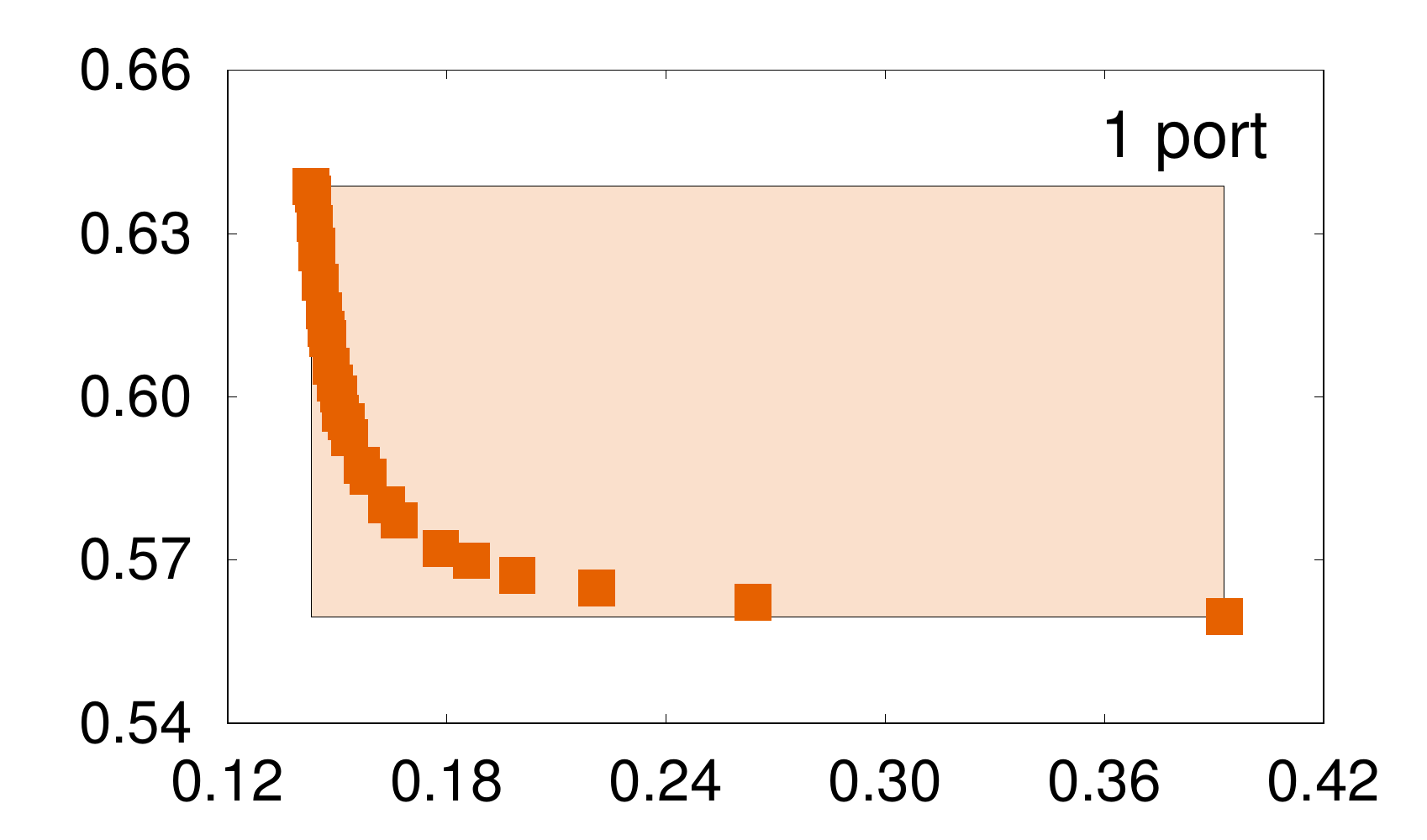} &
\includegraphics[width=0.25\linewidth]{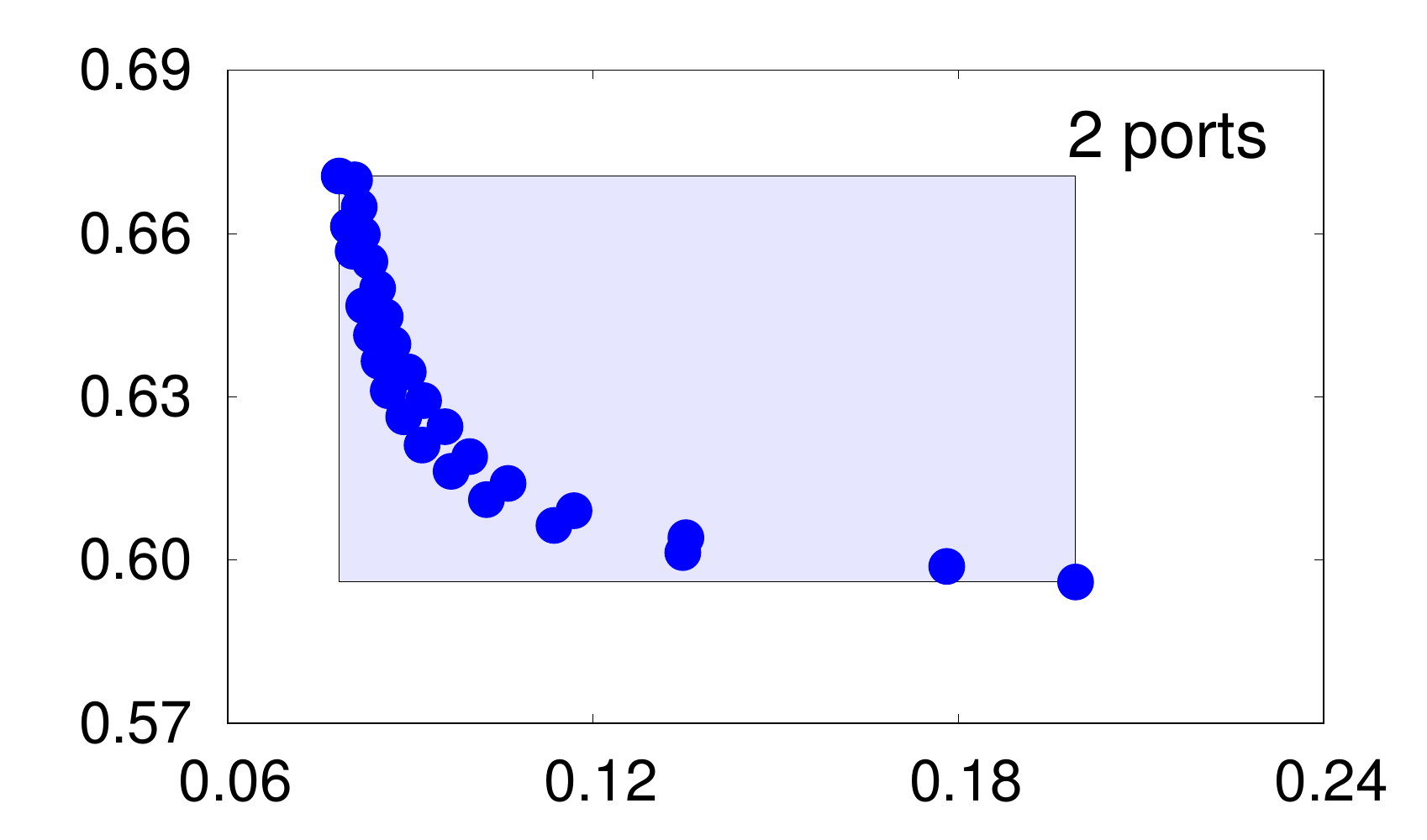} \\
\includegraphics[width=0.25\linewidth]{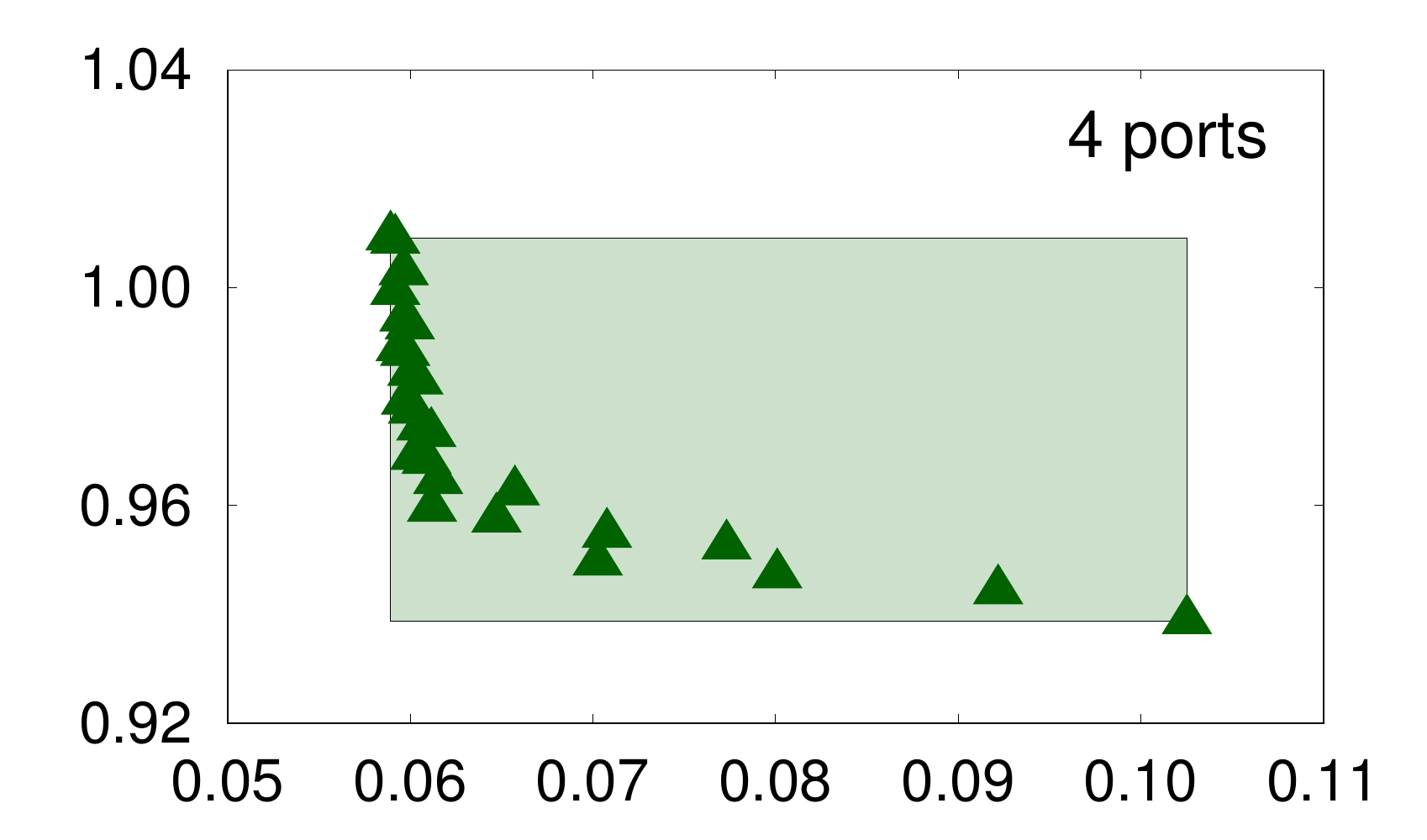} &
\includegraphics[width=0.25\linewidth]{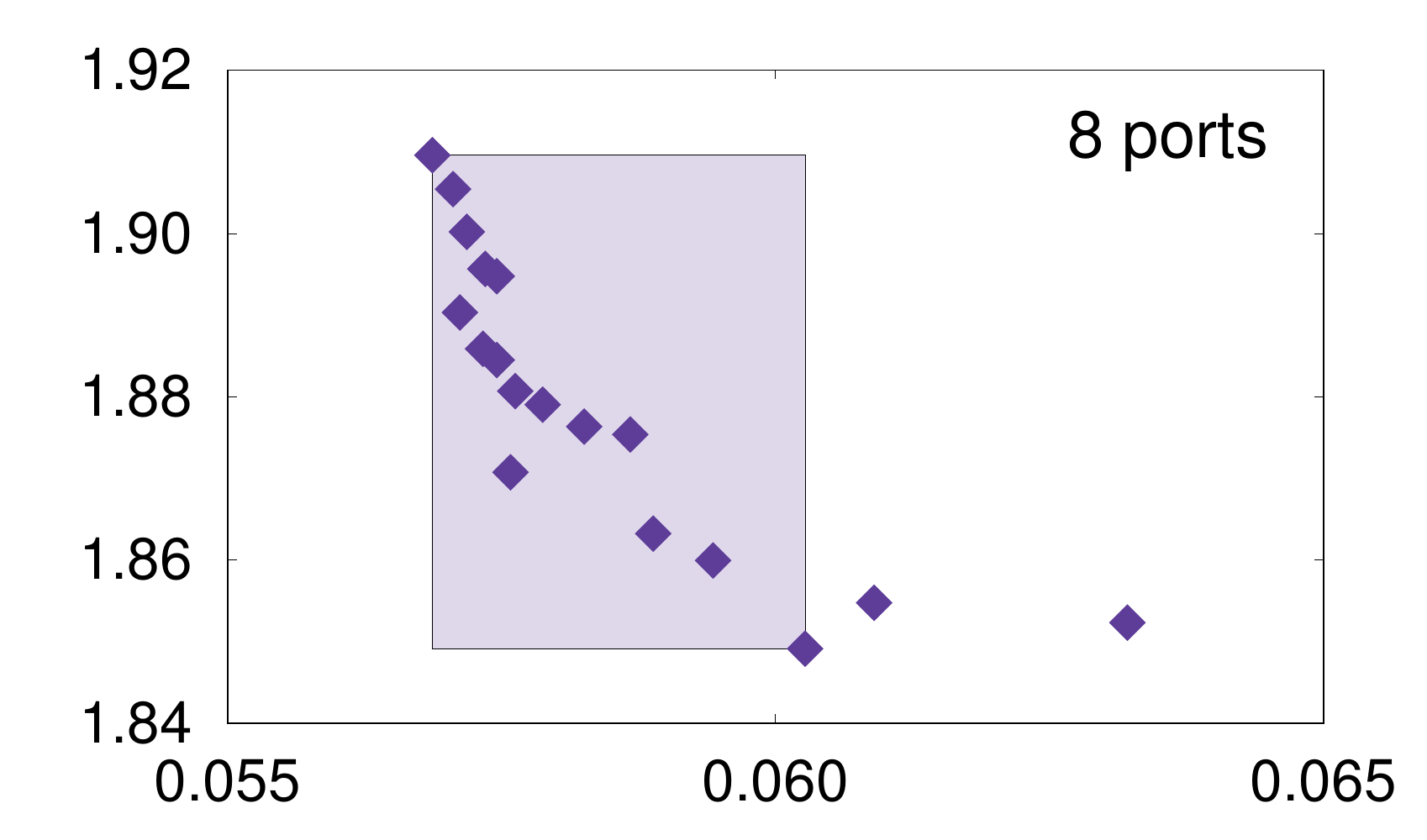} \\
\end{tabular}}
\end{tabular}}\\
\vspace{-0.3cm}%
\subfigure[{\sc\normalsize Change-Det}]{
\begin{tabular}{c@{\hspace{0.2cm}}c}
\imagetop{\includegraphics[width=0.56\linewidth]{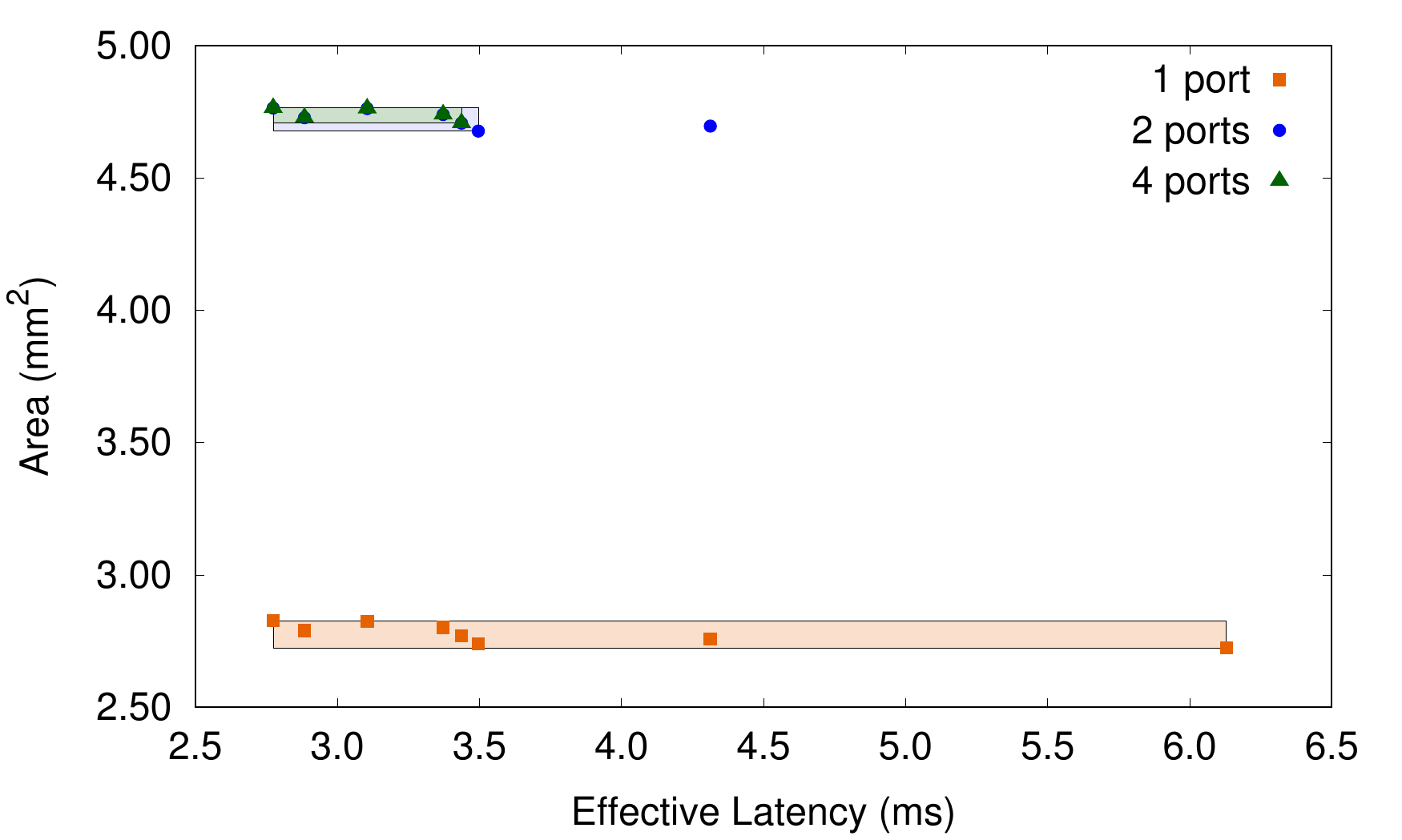}} &
\imagetop{\begin{tabular}{c@{\hspace{0.2cm}}c}
\includegraphics[width=0.25\linewidth]{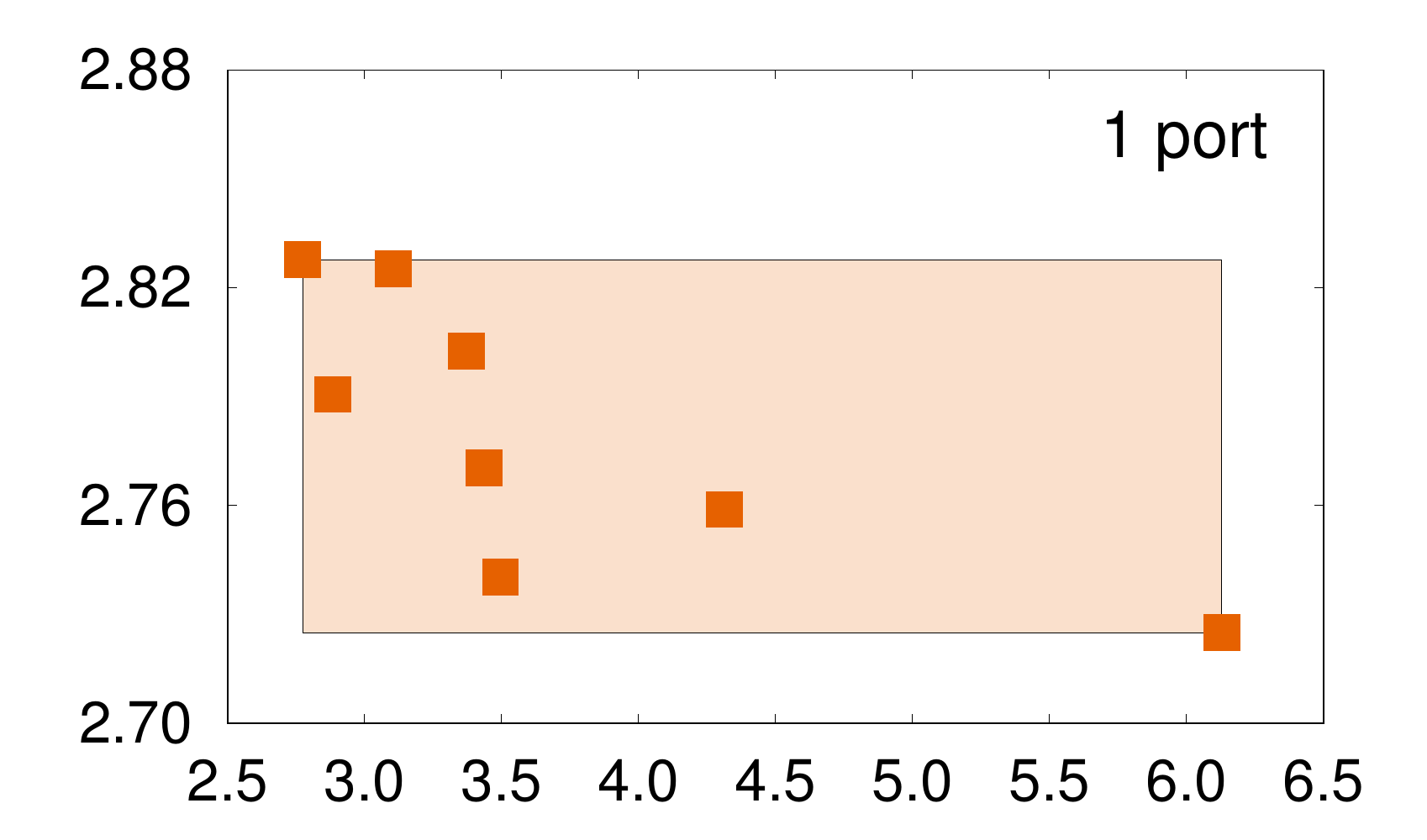} &
\includegraphics[width=0.25\linewidth]{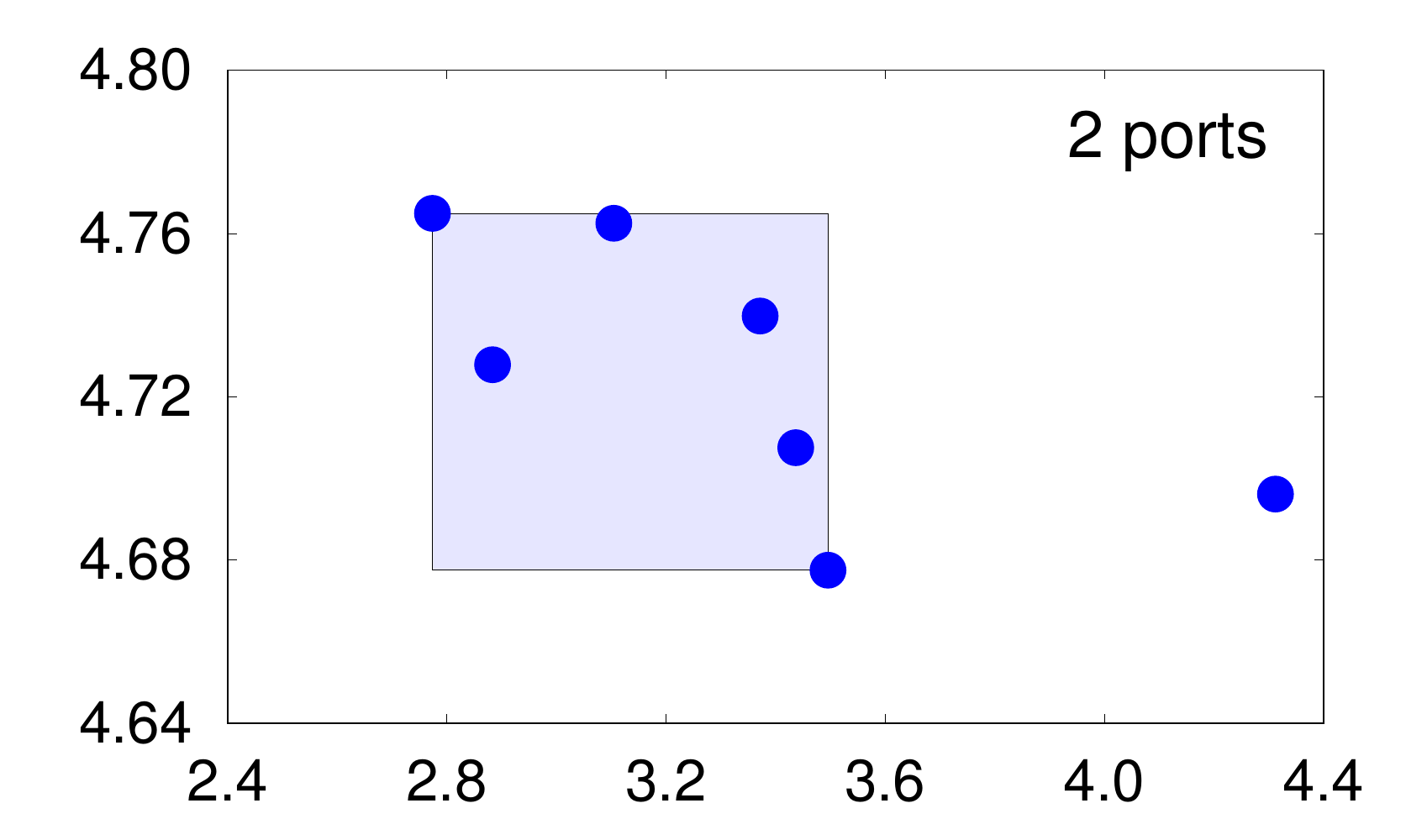} \\
\includegraphics[width=0.25\linewidth]{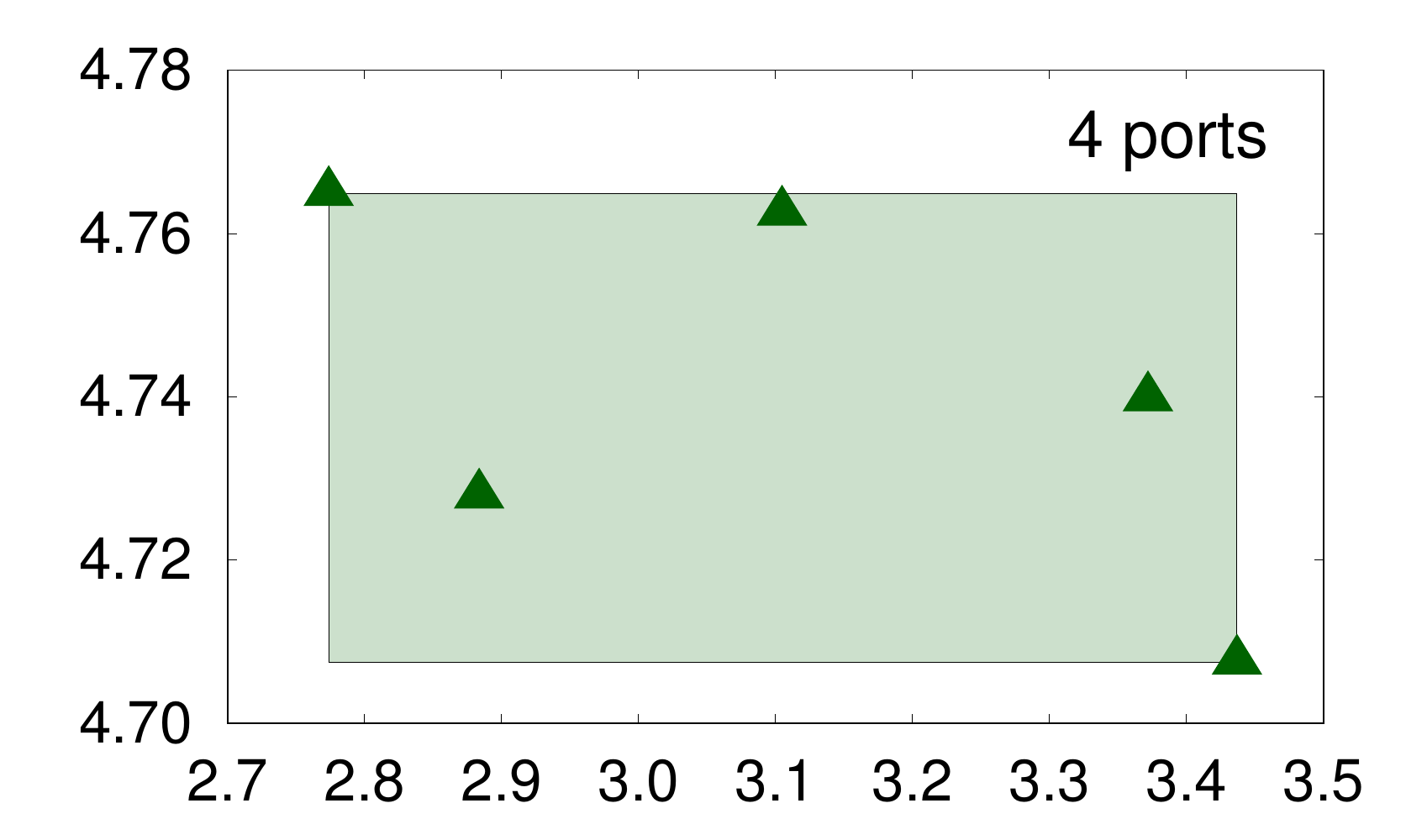} \\
\end{tabular}}
\end{tabular}}
\end{tabular}}
\caption{Characterization of four representative components of the WAMI accelerator.}\label{figure:characterization}
\end{figure*}

\subsection{Design-Space Exploration}

\begin{figure}
\centering\resizebox{0.75\linewidth}{!}{
\includegraphics{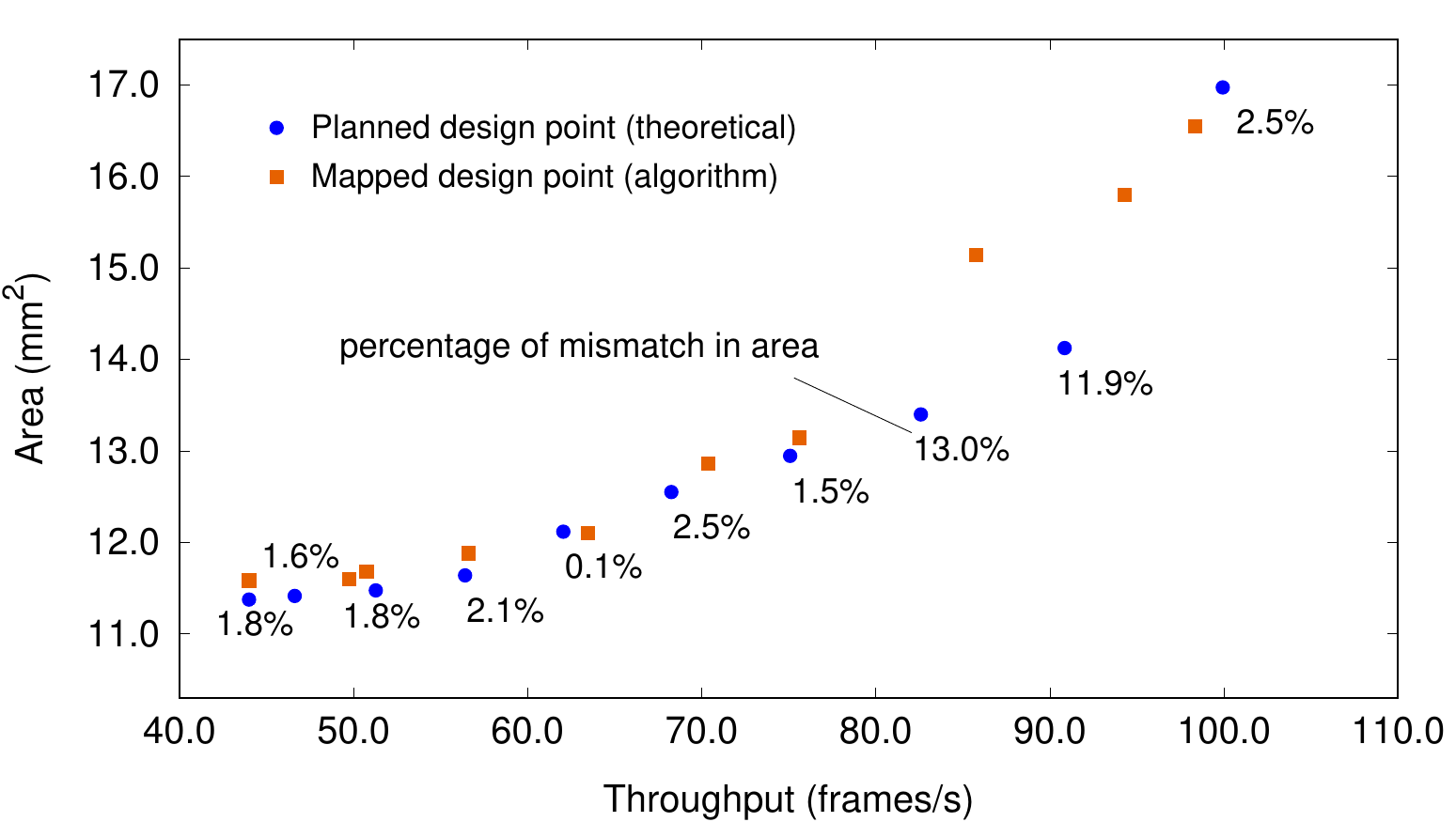}}
\caption{Results of the compositional DSE for WAMI.}\label{figure:system}
\end{figure}

After the characterization of the single components, \toolname applies the DSE
approach explained in Section~\ref{section:system}. It first finds the optimal
solutions at the system level by using Equation~(\ref{equation:min_alpha})
(Section~\ref{section:system:planning}). It then applies the mapping function
to determine the knob settings of the single components and runs the necessary
synthesis (Section~\ref{section:system:mapping}).
\figurename~\ref{figure:system} shows the resulting Pareto curve that includes
the planned points (from Equation~(\ref{equation:min_alpha})) and the mapped
points (returned by the mapping function). These design points
are characterized in terms of effective throughput
(\textit{frame/s}) and area (\textit{mm}$^2$). To quantify the mismatch between
the planned points and the mapped points we calculate the following ratio: 
$$
\sigma(d_p, d_m) = \frac{\mid d_m - d_p \mid}{d_p}
$$
where $d_p$ is the area of a planned point $p$, while $d_m$ is the area of
the corresponding mapped point $m$. Each planned point in
\figurename~\ref{figure:system} is labeled with its corresponding $\sigma\%$ value.
Note that the curve obtained with LP is a theoretical curve because the points
found at the system level do not guarantee the existence of a corresponding set
of implementations for the components. The error is mainly due to the impact of
the memory, which determines a significant distance between two consecutive
regions (e.g., the points with more than 10\% of mismatch in \figurename~\ref{figure:system}). In fact, if a point is mapped between two regions it must be approximated
with the lower-right point of the next region with lower effective latency. This choice permits to
satisfy the throughput requirements almost always, but at the expense of additional area.
In fact, even if Equation~(\ref{equation:min_alpha}) is constrained by the system throughput,
it is not always guaranteed to obtain the same throughput because it is not always the case
that there exists a mapped point that has exactly the same latency of a planned
point. To solve this issue, one could try to reduce the clock period and satisfy
the throughput requirements.  

{
Finally, to demonstrate the efficiency of \toolname,
\figurename~\ref{figure:invocations} shows the number of invocations to the HLS tool.
For each component of WAMI, the right bars report the breakdown of the synthesis
calls performed in each phase of the algorithm. At least two invocations are
necessary for each region to characterize a component. Then, we have to
consider the invocations that fail due to the $\lambda-constraints$, and
finally, the invocations required at system level on the most critical components
(mapping). Some components do non play any role in the efficiency of
the system. For example, for \accsmall{Matrix-Mul}, there are no invocations after
the characterization because only the slowest version has been requested by 
Equation~(\ref{equation:min_alpha}) (to save area). This component is not important to
guarantee a high throughput for the entire system. Moreover, some synthesized
points belong to multiple solutions of the LP problem, as in the case of
\accsmall{Debayer}. Therefore, \toolname avoids performing an invocation of
the HLS with the same knobs more than once. On the other hand, the left bars in
\figurename~\ref{figure:invocations} report the number of invocations required for a exhaustive
exploration. Such exploration requires to (i) synthesize
all the possible configurations of unrolls and memory ports for each
component, (ii) find the Pareto-optimal design points for each component, and (iii) compose all the Pareto-optimal designs to find the
Pareto curve at the system level (Section~\ref{section:example}).  The left bars in
\figurename~\ref{figure:invocations} show the number of invocations to the HLS tool required in step (i). \toolname
reduces the total number of invocations for WAMI by
$6.7\times$ on average and up to 14.6$\times$ for the single components,
compared to the exhaustive exploration. {Further, while \toolname returns the Pareto-optimal
implementations at the system level, to find the combinations of the components that are Pareto
optimal with an exhaustive search method, one has to combine
the huge number of solutions for the single components.} In the case of WAMI,
the number of combinations, i.e., the product of the number of Pareto-optimal points of
each component, is greater than $9*10^{12}$. This motivates the need of using a compositional
method like \toolname for the DSE of complex accelerators.
\fillparagraph
}


\begin{figure}
\centering\resizebox{0.58\linewidth}{!}{
\includegraphics{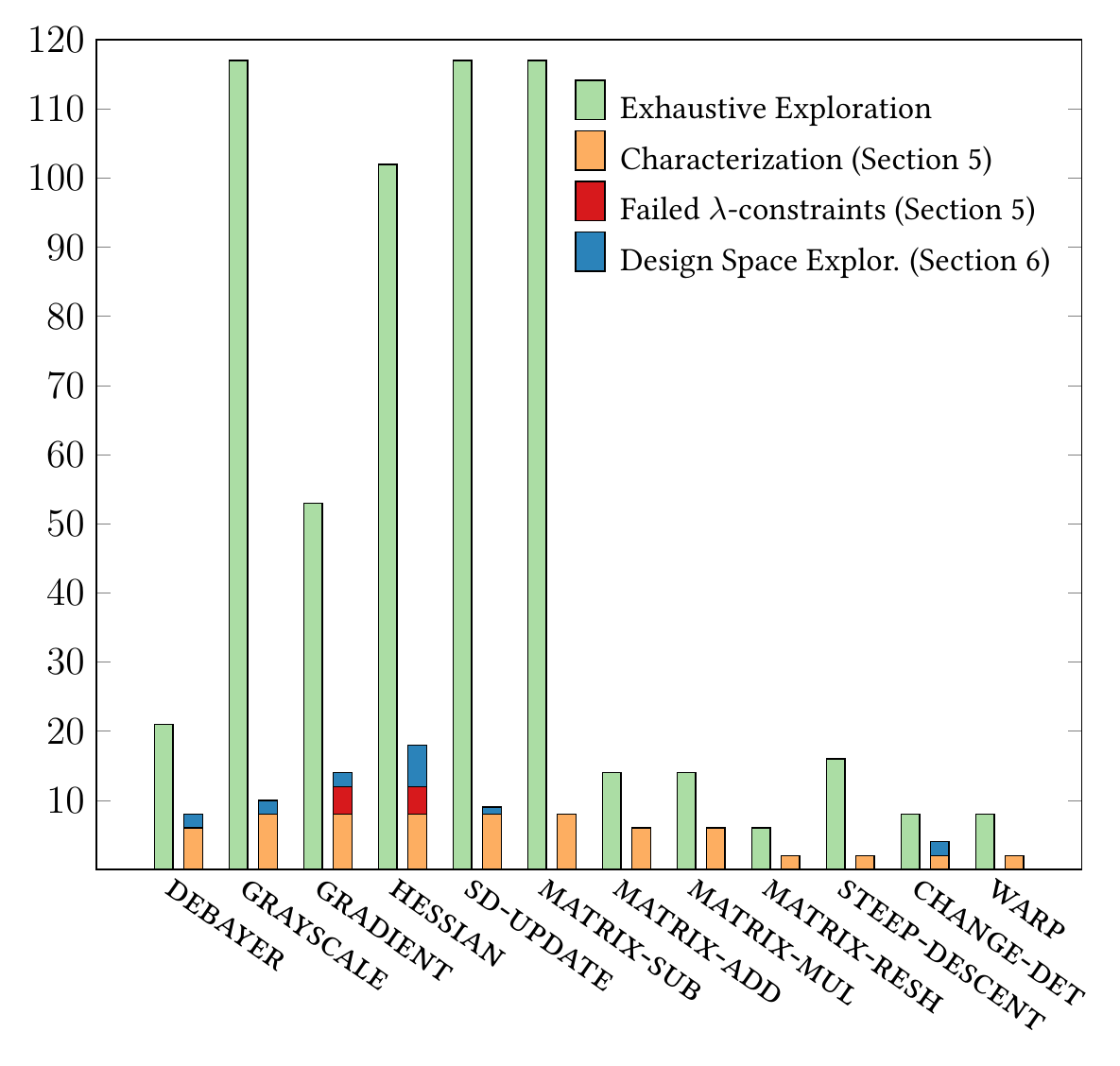}}
\caption{Number of invocations of the HLS tool for an exhaustive exploration (bars on the left) and \toolname (on the right). }\label{figure:invocations}
\end{figure}

\subsection{Summary}

We report a brief summary of the achieved results:

\begin{itemize}
\itemsep 0.1em
\item \toolname guarantees a \textbf{richer} DSE with respect to the approaches
that do not consider the memory as integral part of the DSE: for WAMI, 
\toolname guarantees an average performance span of $4.06\times$ and
an average area span of $2.58\times$ as opposed to $1.73\times$ and
$1.22\times$, respectively, when only standard dual-port memories are
used; {\toolname obtains a richer set of Pareto-optimal
implementations thanks to memory generation and optimization;}
\item \toolname guarantees a \textbf{faster} DSE compared to exhaustive search
methods: for WAMI, \toolname reduces the number of invocations 
to the HLS tool by $6.7\times$ on average and by up to $14.6\times$ for the single
components; {\toolname is able to reduce the number of invocations thanks
the compositional approach discussed in Section~\ref{section:system}};
\item \toolname is an \textbf{automatic} and \textbf{scalable} methodology for DSE: the approach is 
intrinsically compositional, and thus with larger designs the performance gains are expected 
to be as good as smaller ones, if not better. While an exhaustive method has to explore all 
the alternatives, \mbox{\toolname focuses on the most critical components.}
\end{itemize}

\section{Related Work}\label{section:related}

{
This section describes the most-closely related methods to perform DSE. We
distinguish the methods that explore single-component designs (reported in
Section~\ref{section:related:comp}) from those that are compositional like
\toolname (in Section~\ref{section:related:syst}). 
}

%
%

\subsection{Component DSE}\label{section:related:comp}

{
Several methods have been proposed to drive HLS tools for DSE. There exist
probabilistic approaches~\cite{schafer_tcad16}, search algorithms based on
heuristics, such as simulated annealing~\cite{schafer09}, iterative
methodologies that exploit particle-swarm optimization~\cite{mishra_isqed14},
as well as genetic algorithms~\cite{ferrandi_svlsi08}, and
machine-learning-based exploration methodologies~\cite{mahapatra_eslsyn14,
liu_dac13, schafer_iet12}. All these methods try to quickly predict the
relevance of the knobs and determine the Pareto curves of the scheduled RTL
implementations in a multi-objective design space.  None of these methods,
however, consider the generation of optimized memory subsystems for hardware
accelerators. Conversely, other methods focus on creating efficient memory
subsystems, but without exploring the other HLS knobs. For instance, Pilato et
al.~\cite{pilato_codes14} propose a methodology to create optimized memory
architectures, partially addressing the limitations of current HLS tools in
handling memory subsystems. This enables a DSE that takes into account also the
memory of accelerators. However, that work focuses on optimizing the memory
architectures and not in proposing efficient DSE methods. Similarly, Cong et
al.~\cite{cong_tcad2016} explore memory reuse and non-uniform partitioning for
minimizing the number of banks in multi-bank memory architectures for stencil
computations. Differently from these works, \toolname coordinates both memory
generators, like the one proposed in~\cite{pilato_tcad17}, and HLS tools to
find several Pareto-optimal implementations of accelerators.  Other
methodologies apply both loop manipulations and memory optimizations. For
instance, Cong et al.~\cite{cong_dac2012,cong_iccad2011} adopt polyhedral-based
analysis to apply loop transformations with the aim of optimizing memory reuse
or partitioning.  Differently from these works, \toolname focuses on
configuring the knobs provided by HLS, after applying such loop
transformations.  Indeed, \toolname realizes a compositional-based methodology,
and thus it finds Pareto-optimal implementations of the entire system, and not
only of the single components.  The first step of \toolname consists in the
characterization of components to identify regions of the multi-objective
design space where feasible RTL implementations exist.  This step differs from
previous works~\cite{schafer_tcad16,caps_dac11,liu_date12}  for two main
aspects. First, \toolname includes memory generation and optimization in the
DSE process. Second, \toolname applies synthesis constraints to account for the
high variability and partial unpredictability of the HLS tools. Such
constraints consider both the dependency graph of the specification and the
memory references in each loop.  Thus, \toolname identifies larger regions of
Pareto-optimal implementations.
\fillparagraph
}

Other methods, such as \emph{Aladdin}~\cite{shao2014}, perform a DSE without
using HLS tools and without generating the RTL implementations, estimating the
performance and costs of high-level specifications (C code for \emph{Aladdin}).
\toolname differs from these methods because it aims at generating efficient
RTL implementations by using HLS and memory generator tools. Indeed, such
methods can be used \emph{before} applying \toolname to pre-characterize the
different components of an accelerator that is not ready to be synthesized with
HLS tools. Since the design of HLS-ready specifications requires significant
efforts~\cite{qamar_17}, this can help the designers to focus only on the most
critical components, i.e., those that are expected to return good performance
gains over software executions. After this pre-characterization, \toolname can
be used to perform a DSE of such components and obtain the Pareto-optimal
combinations of their RTL implementations.

%
%

\subsection{System DSE}\label{section:related:syst}

While the previous approaches obtain Pareto curves for single components, only
few methodologies adopt compositional design methods for the synthesis of
complex accelerators. The approach used by \toolname predicts the Pareto curve
at the system level, similarly to those proposed by Liu et
al.~\cite{liu_date12} and Haubelt and Teich~\cite{haubelt_aspdac03}.
Differently from these works, \toolname correlates also the planned design
points, which are simply theoretical (the LP solutions), with feasible
high-level knob settings and memory configuration parameters.  Further,
\toolname focuses on optimizing the HLS knobs, e.g., loop manipulations, and
memory subsystems, rather than tuning low-level knobs, e.g., the target clock
period.


%
%

\section{Concluding Remarks}\label{section:concl}

We presented \toolname, an automatic methodology for compositional DSE that
coordinates both HLS and memory generator tools. \toolname takes into account
the unpredictability of the current HLS tools and considers the PLMs of the
components as an essential part of the DSE. The methodology of \toolname is
intrinsically compositional. First, it characterizes the components to define
the regions of the design space that contain Pareto-optimal implementations.
Then, it exploits a LP formulation to find the Pareto-optimal solutions at the
system level. Finally, it identifies the knobs for each component that can be
used to obtain the corresponding implementations at RTL.  We showed the
effectiveness and efficiency of \toolname by considering the WAMI accelerator
as a case study. Compared to methods that do not consider the PLMs, \toolname
finds a larger set of Pareto-optimal implementations.  Additionally, compared
to exhaustive search methods, \toolname reduces the number of invocations to
the HLS tool by up to one order of magnitude.

%
%

\begin{acks}
The authors would like to thank the anonymous reviewers for their valuable comments and helpful 
suggestions that help us improve the paper considerably. This work was supported in part by DARPA 
PERFECT (C\#:  R0011-13-C-0003), the National Science Foundation (A\#: 1527821), and C-FAR 
(C\#: 2013-MA-2384), one of the six centers of STARnet, a Semiconductor Research 
Corporation program sponsored by MARCO and DARPA.
\end{acks}

%
%

\bibliographystyle{ACM-Reference-Format}
\bibliography{references} 

\end{document}